\documentclass{article}
\usepackage{graphicx} 
\usepackage{natbib}
\usepackage{url}
\usepackage{amsmath, mathtools, amsthm, amssymb, mathrsfs}
\usepackage{geometry}
\usepackage{comment}
\usepackage{subfig}
\usepackage[dvipsnames]{xcolor}
\usepackage{algorithm}
\usepackage{algpseudocode}
\usepackage{hyperref}
\usepackage{cleveref}
\hypersetup{
    colorlinks=true, 
    linkcolor=blue, 
    urlcolor=black, 
    citecolor=blue
}

\title{Revisiting Penalized Likelihood Estimation for Deterministic Computer Experiments}
\author{Ayumi Mutoh\thanks{Corresponding author: Department of Statistics, NC State University, {\tt amutoh@ncsu.edu} }
\and Annie S. Booth\thanks{Department of Statistics, Virginia Tech}
\and Jonathan W. Stallrich\thanks{Department of Statistics, NC State University}}

\date{\today}

\begin{document}

\maketitle

\begin{abstract}
Gaussian processes (GPs) are popular as nonlinear regression models for expensive computer simulations, yet GP performance relies heavily on estimation of unknown covariance parameters. Maximum likelihood estimation (MLE) is common, but it can be plagued by numerical issues in small data settings.  The addition of a nugget may help but is not a cure-all.  Penalized likelihood methods may improve upon traditional MLE, but their success depends on tuning parameter selection.  We introduce a new cross-validation (CV) metric called ``decorrelated prediction error'' (DPE), within the penalized likelihood framework for GPs. Inspired by the Mahalanobis distance, DPE provides more consistent and reliable tuning parameter selection than traditional metrics like prediction error, particularly for $K$-fold CV. Our proposed metric performs comparably to standard MLE when penalization is unnecessary and outperforms traditional tuning parameter selection metrics in scenarios where regularization is beneficial, especially under the one-standard error rule. 
\end{abstract}


\section{Introduction}\label{sec:intro}
Computer experiments (or ``simulators'') are valuable tools in the study of complex physical processes, offering an alternative to expensive or time-consuming physical experiments.  Applications span engineering \citep{chen2006compexpretiment} including automotive crash analysis  \citep{berthelson_auto}, building design \citep{westermann_building}, and the study of natural hazards \citep{bayarri_volcano}.  The computational complexity of computer experiments often limits the number of times the simulator can be evaluated in practice.  Limited simulation data necessitates a surrogate model (or ``emulator'') to stand-in-place of true simulator evaluations at unobserved input settings. We will focus on simulators that produce a deterministic output $y(\mathbf{x})$ from $d$ scaled input settings, $\mathbf{x} \in [0,1]^d$. A good surrogate should be flexible, interpretable, and quick to evaluate.  Most importantly, it should provide accurate predictions with appropriate uncertainty quantification (UQ) across the entire input space, including the training data locations which should have no uncertainty. This can be a tall order when training data is scarce, which is common for computationally complex simulators.

While a variety of surrogate modeling approaches exist \citep[e.g.,][]{chen2006compexpretiment,levy2010,kudela2022}, Gaussian processes (GPs) are at the forefront \citep{sacks1989,Santner2003,gramacy2020}.  GP surrogates are based on tractable conditional Normal distributions, with predicted values equal to the conditional GP mean. The covariance function is the workhorse of any GP surrogate.  It specifies the relationship between response values as a function of their input locations.  Typical covariance functions are inverse functions of Euclidean distance, further parameterized by a scale (or ``variance'') and lengthscale. We will work with the popular squared exponential covariance function with scale parameter $\sigma^2$, lengthscales $\theta_p \geq 0$, and nugget $g$
\begin{align}
\Sigma(\mathbf{x}_i,\mathbf{x}_j)=\sigma^2 R(\mathbf{x}_i,\mathbf{x}_j)=\sigma^2 \left( \exp\left(-\sum_{p=1}^d\theta_p (x_{ip}-x_{jp})^2\right)+g \mathbb{I}_{(i=j)}\right)\ ,\ \label{eqn:cov_func}
\end{align}
which is defined on input settings $\mathbf{x}_i$ and $\mathbf{x}_j$ from the training data.\footnote{It is also common to use the reciprocal of $\theta_p$ here, but that parameterization is incompatible with penalization.  In our parameterization, penalties encourage smaller $\theta$ and discourage abrupt correlation decay.}  The nugget $g \geq 0$ corresponds to the given simulator's variability at a fixed $\mathbf{x}$, which only appears when $i=j$. Given that we are focused on deterministic simulators, it seems appropriate to set $g=0$, but we will soon argue how including a small, fixed $g$ can provide near interpolation at the training data along with improved UQ. Proper estimation of the $\theta_p$'s can make-or-break a GP surrogate. If they are estimated to be too small or too large, the surrogate will be prone to underestimate or overestimate uncertainty, respectively.

Limited sample sizes are common when dealing with expensive computer experiments. In such settings, several approaches have been proposed to address the resulting challenges in GP covariance estimation. With few observations, the empirical covariance matrix can become ill-conditioned, leading to unstable lengthscale estimates and unreliable uncertainty quantification. Low rank covariance regularization mitigates these issues by approximating the full covariance matrix using a reduced rank structure and has been widely applied in small data settings \citep{Rasmussen2006, quinonero2005unifying, titsias2009variational}. In contrast, our work focuses instead on direct parameter estimation rather than covariance approximation.

While cross validation (CV) methods and Bayesian posterior sampling are possible \citep[e.g.,][]{geisser1979cv,mackay1992bayes},
maximum likelihood estimation (MLE) of covariance hyperparameters is usually favored for its simplicity and computational efficiency.  As evidence of MLE's popularity, most popular GP software packages adopt MLE as their default estimation method.  These include: \texttt{GPML} \citep{Rasmussen2010gpml} and \texttt{DACE} \citep{neilsen2002DACE} in MATLAB; \texttt{Scikit-learn} \citep{pedregosa2011scikitlearn}, \texttt{GPy} \citep{gpy2014}, and \texttt{GPflow} \citep{matthews2016} in Python; and \texttt{GPfit} \citep{GPfit2015}, \texttt{DiceKriging} \citep{dicekriging}, and \texttt{mlegp} \citep{mlegp} in R. Unfortunately, numerical convergence issues can arise when maximizing the GP likelihood, especially with small data sizes.
Tricks to circumvent these issues in the GP likelihood include reparameterizing the covariance function or bounding the possible values of the parameters \citep[e.g.,][]{GPfit2015, butler2014likelihood, basak2021issues, binois2021hetgp}. If convergence issues in the GP likelihood stem from a nearly-singular covariance matrix, they may be allayed by the addition of a $g>0$ \citep{gramacy2012nugget, andrianakis2012effect}, although this comes at the cost of not interpolating the training data. 

\begin{figure}[!ht]
    \centering
    \subfloat[Profile log likelihoods\label{fig:1a}]{\includegraphics[width=0.3\textwidth, height=0.27\linewidth]{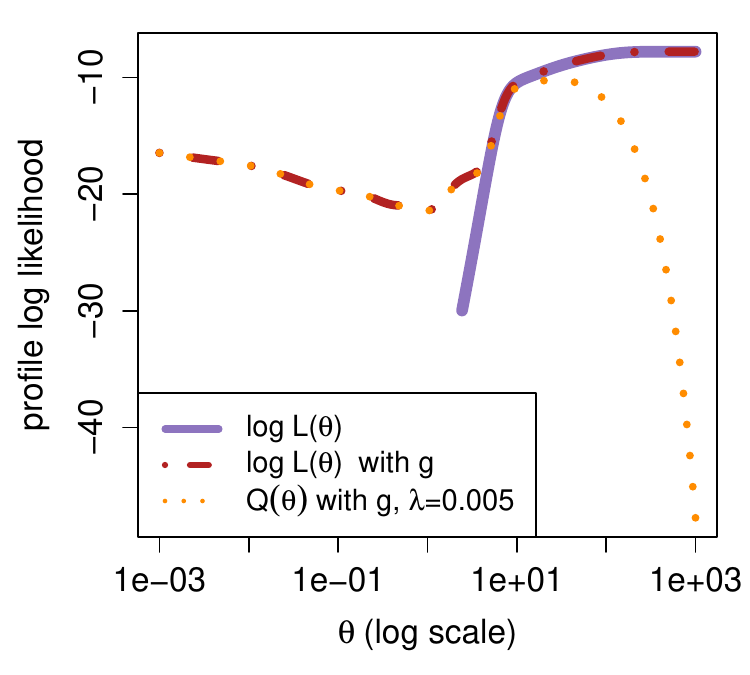}}
     \subfloat[$\hat{\theta}=1000$ with fixed g\label{fig:1b}]{ \includegraphics[width=0.3\textwidth, height=0.27\linewidth]{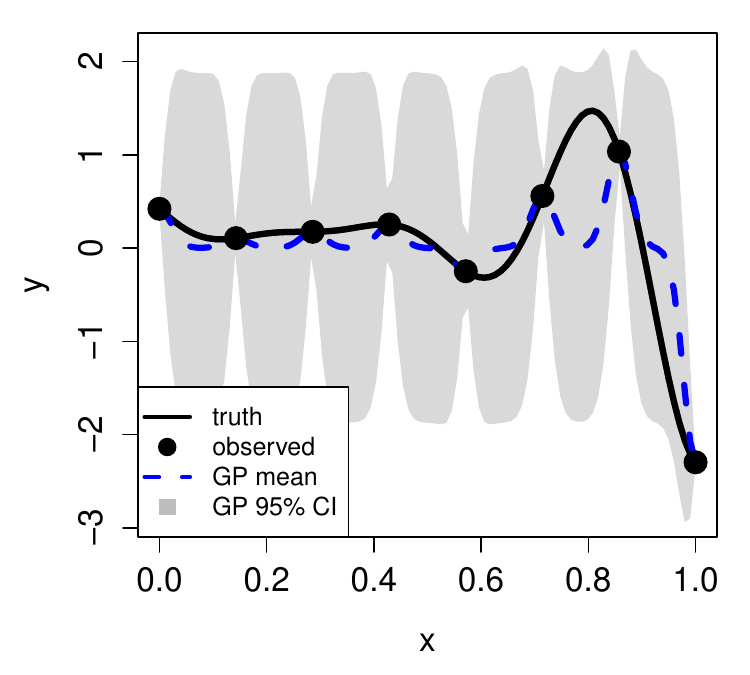}}
 \subfloat[$\hat{\theta}=922.203$ with $\hat{g}$\label{fig:1c}]{\includegraphics[width=0.3\textwidth, height=0.27\linewidth]{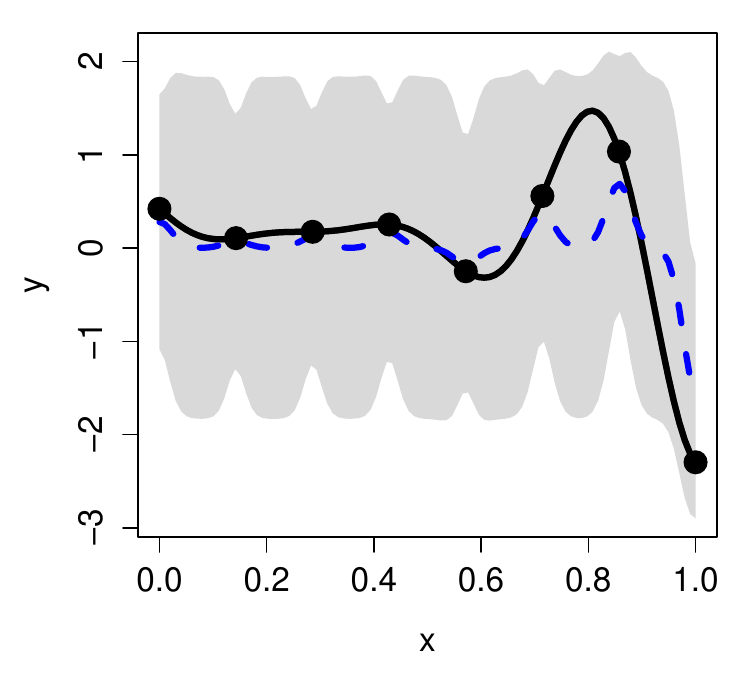}} \\
 \subfloat[$\hat{\theta}=27.444$ with fixed g\label{fig:1d}]{ \includegraphics[width=0.3\textwidth, height=0.27\linewidth]{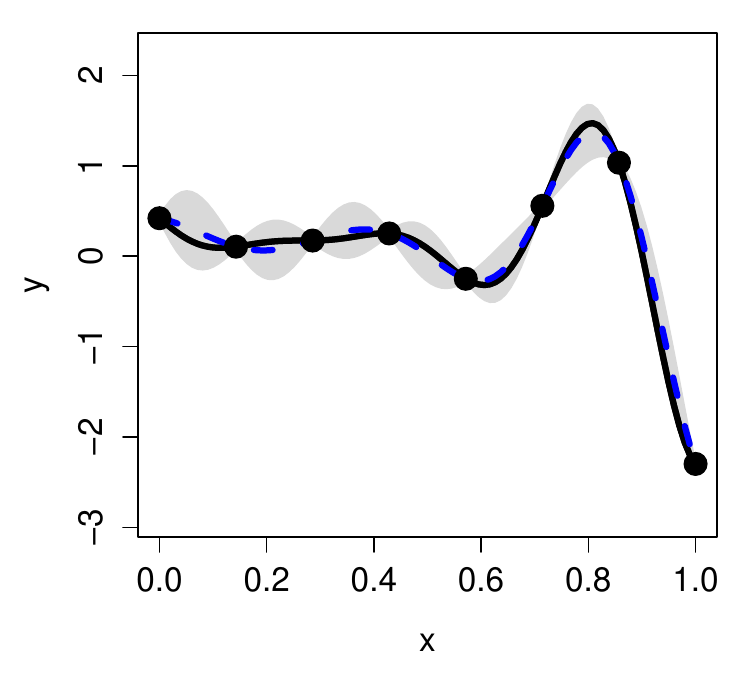}} 
    \subfloat[$\hat{\theta}=10.565$ with fixed g\label{fig:1e}]{\includegraphics[width=0.3\textwidth, height=0.27\linewidth]{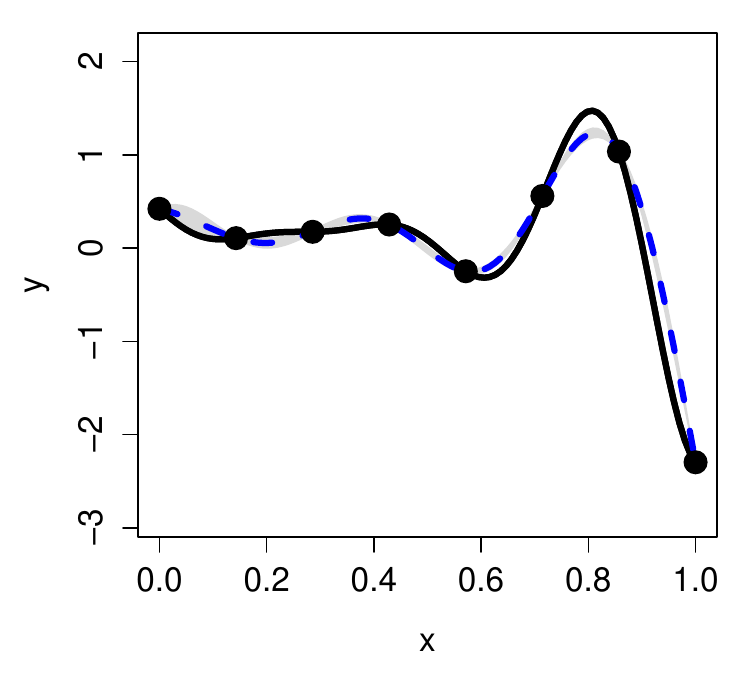}}
    \subfloat[$\hat{\theta} = 0.001$ with fixed $g$\label{fig:1f}]{\includegraphics[width=0.3\linewidth, height=0.27\linewidth]{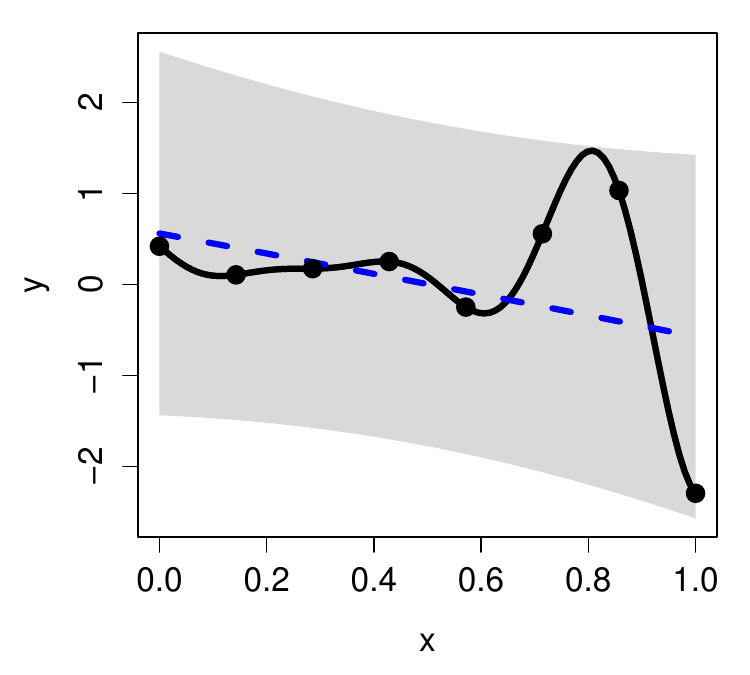}}
    \caption{(a) Various profile log likelihoods for the Forrester example.  
    All curves are plotted over $\theta\in[0.001, 1000]$ on the log scale. (b) GP surrogate using the MLE $\hat{\theta}$ from $\log\mathcal{L}(\theta)$ with $g=10^{-5}$. (c) GP surrogate using the jointly estimated MLEs $\hat{\theta}$ and $\hat{g}$. (d) GP surrogate using $\hat{\theta}=27.444$, the maximizer of $Q(\theta)$ with $\lambda=0.005$. (e) GP surrogate constracted using the $\hat{\theta}$ selected via $n$-fold CV on prediction error.  (f) GP surrogate constructed using the $\hat{\theta}$ selected via $n$-fold CV on prediction error, with 1SE rule applied. 
    }
    \label{fig:forrester}
\end{figure}

To demonstrate these issues, we present a toy example which utilizes GP fundamentals reviewed in Section~\ref{sec:review}. We consider the one-dimensional Forrester function \citep{forrester2008}, $y=(6x-2)^2\sin{(12x-4)}$, with $n=8$ equally spaced data points over the original domain $x\in[0,1.25]$ (with inputs scaled to the unit interval, which we will do throughout). For $g=0$, the corresponding profile log likelihood, $\log \mathcal{L}(\theta)$ is shown by the purple solid line in \Cref{fig:1a} with $\theta$ on the log scale). As $\theta \to 0$, $\log \mathcal{L}(\theta)$ disappears due to numerical issues with matrix inversion. The dark red dot-dashed line in \Cref{fig:1a} corresponds to $\log \mathcal{L}(\theta)$ with $g=10^{-5}$, which is now well-defined for all $\theta$. But another issue arises: the MLE for $\theta$ is 1000, producing a practically useless surrogate due to poor predictions outside the training data settings and excessively large prediction intervals as shown in \Cref{fig:1b}. Estimating $g$, as advocated by \cite{gramacy2012nugget}, slightly reigns in the MLE for $\theta$, but its surrogate, shown in \Cref{fig:1c}, has the same issues as in \Cref{fig:1b} and even fails to perform well at the training data locations. Because our focus is on deterministic computer experiments, where the GP surrogate is expected to interpolate the data, we henceforth fix $g=10^{-5}$.

Penalized likelihood estimation is a common tool to reign in the log likelihoods like that shown in \Cref{fig:1a} \citep{Tibshirani1996lasso, ridge1970, fu1998bridge}, but it has not received the same attention for GPs.  \citet{li2005} proposed a penalized log likelihood, $Q(\theta)$, to discourage extremely large lengthscale estimates via a penalty $p_\lambda(\theta)$ on the magnitude of $\theta$.  The penalty includes a tuning parameter, $\lambda \geq0$, which influences the importance of the penalty relative to $\log \mathcal{L}(\theta)$. Setting $\lambda=0$ removes the penalty and $\lambda \to \infty$ makes $\theta \to 0$. The orange dotted line in \Cref{fig:1a} demonstrates the penalized log likelihood with the LASSO penalty for a fixed $g$ and $\lambda=0.005$. The corresponding surrogate shown \Cref{fig:1d} is definitely superior to the two unpenalized competitors, having both superior predictions and UQ.

Given the potential of penalized likelihood estimation, it is surprising the method has been largely ignored in the computer experiments literature.  For example, \cite{roustant2012} investigate the penalized GP estimation included in \texttt{DiceKriging} with disappointing results. After performing our own investigation, we identified tuning parameter selection, a necessary component of penalized estimation, as the main limitation. \cite{li2005} proposed selecting $\lambda$ by minimizing the surrogate's prediction error via $K$-fold CV. We performed $n$-fold, or leave-one-out CV, on the Forrester example and identified the $\lambda$ that minimized prediction error. The resulting surrogate, shown in \Cref{fig:1e}, shows improved predictions compared to the unpenalized competitors, but poor coverage for larger values of $x$.

\begin{figure}[!ht]
    \centering
    \includegraphics[height=0.31\linewidth]{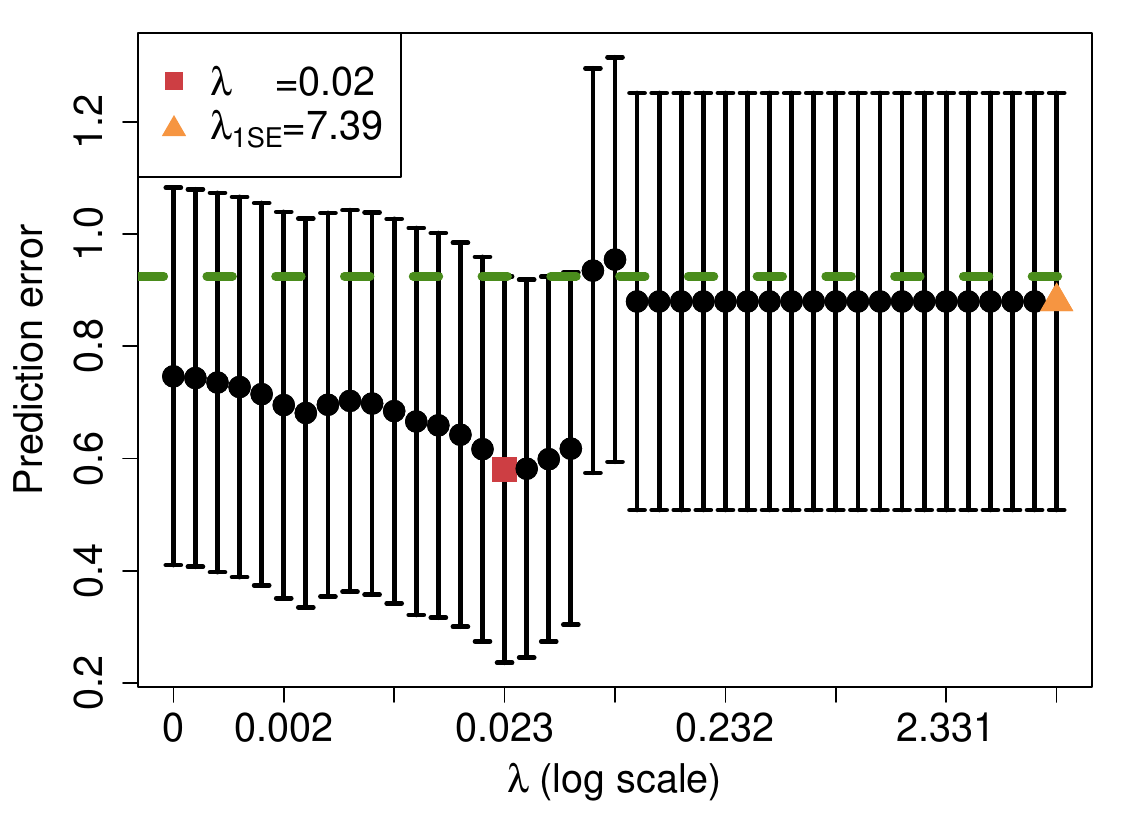}
    \caption{Average $n$-fold prediction error for the Forrester function across $\lambda \in [0,7.39]$. Black vertical lines indicate plus/minus one standard error.  Red square marks the $\lambda$ values that minimize average prediction error ($\lambda = 0.02$).  Orange triangle marks the largest $\lambda$ within one standard error of the ``best'' $\lambda$ ($\lambda_{\textrm{1SE}}=7.39$).
    } 
    \label{fig:figure2}
\end{figure}

Another common approach for selecting $\lambda$ is the ``one-standard-error'' (1SE) rule \citep{breiman1984onese, chen2021onese}.  This involves selecting the smallest $\lambda$ value whose average performance metric is within one standard error of the actual lowest performance metric. \Cref{fig:figure2} shows the mean prediction errors and their standard errors across $\lambda\in[0,7.39]$ for the Forrester example. The red square marks the degree of penalization that resulted in the lowest mean prediction error, which selected $\lambda=0.02$. The green dashed line marks one standard error above the lowest mean prediction error, and the orange triangle marks the largest $\lambda$ value whose mean prediction error is below that threshold. The significant variability across the leave-one-out folds leads to even more aggressive penalization.  The resulting surrogate, shown in \Cref{fig:1f}, has over-smoothed predictions that fail to interpolate the training data and has large UQ. The punchline: penalization under some $\lambda$ can improve upon MLE, but identifying such $\lambda$ is challenging for small data sets.

The main contribution of this paper is a new tuning parameter selection metric for penalized likelihood estimation with GP surrogates, called decorrelated prediction error (DPE), which is essentially an external estimator of the scale parameter.  
We compare DPE to prediction error, Mahalanobis distance, and Score, finding that these latter metrics suffer from large between-fold variation for small training data. DPE is shown to exhibit far less variation for interpolating surrogates, and quickly inflates for non-interpolating surrogates, making it a preferable CV metric for tuning parameter selection for deterministic simulators.

The remainder of this article is organized as follows. \Cref{sec:review} reviews GP fundamentals and penalized likelihood estimation. \Cref{sec:method} motivates and describes the DPE metric for $K$-fold CV, including the incorporation of a one-standard-error rule.  It also provides new limiting forms for prediction error, Mahalanobis distance, and Score as $\theta \to 0$ that explain the metrics' potential for poor tuning parameter selection compared to DPE. \Cref{sec:simulation} benchmarks DPE against competing tuning parameter selection strategies on a variety of test functions. \Cref{sec:piston} revisits the Piston slap noise data example from \cite{li2005},
comparing the performance of their $n$-fold CV with prediction error to our $K$-fold CV with DPE. We conclude the paper with a discussion and opportunities for future work in \Cref{sec:discussion}.  All methods, including our proposed DPE metric, are supported by the new {\tt GPpenalty} package on CRAN \citep{mutoh2025package}.

\section{Background}\label{sec:review}
 Let $y_i = f(\mathbf{x}_i)$ denote the $i^\textrm{th}$ observation of a deterministic, real-valued, black-box function $f$ at some $d$-dimensional input $\mathbf{x}_i$. Likewise, let $\mathbf{y}_n = f(\mathbf{X}_n)$ denote the corresponding vector of $n$ observations at row-stacked input locations $\mathbf{X}_n$ of dimension $n\times d$.  A GP prior over $f$ presumes $\mathbf{y}_n\sim\mathcal{N}_n\left(\boldsymbol\mu, \boldsymbol\Sigma_n\right)$.  Prior mean $\boldsymbol\mu$ may be constant or a function of $\mathbf{X}_n$; we will use $\boldsymbol\mu = \mathbf{0}$ after centering responses.  Covariance $\boldsymbol\Sigma_n = \sigma^2 \mathbf{R}_n$ is formed from the ``kernel'' function $R(\mathbf{x}_i, \mathbf{x}_j)$, defined in \eqref{eqn:cov_func} with fixed $g=10^{-5}$, which provides the correlation between $y_i$ and $y_j$. We let $\Omega=\{\boldsymbol{\theta},\sigma^2\}$ denote the full set of unknown GP parameters. Given training data $\mathcal{D}=\left\{\mathbf{X}_n,\mathbf{y}_n\right\}$, predictions for $f$ at $m$ input locations $\mathcal{X}$ follow
\begin{equation}\label{eq:gppost}
\begin{aligned}
    f(\mathcal{X} \mid \mathcal{D}) & \sim \mathcal{N}_m\left( \ \boldsymbol{\mu}_\theta(\mathcal{X} \mid \mathcal{D}),\boldsymbol{\Sigma}_\Omega(\mathcal{X} \mid \mathcal{D}) \ \right)\\
    \boldsymbol{\mu}_\theta(\mathcal{X} \mid \mathcal{D}) & = \mathbf{R}(\mathcal{X}, \mathbf{X}_n) \mathbf{R}_n^{-1}\mathbf{y}_n\\
    \mathbf{\Sigma}_\Omega(\mathcal{X} \mid \mathcal{D}) &= \sigma^2\left( \mathbf{R}(\mathcal{X},\mathcal{X})- \mathbf{R}(\mathcal{X}, \mathbf{X}_n)\mathbf{R}_n^{-1} \mathbf{R}(\mathbf{X}_n,\mathcal{X})\right) \\
    & \equiv \sigma^2 \mathbf{R}_\theta(\mathcal{X} \mid \mathcal{D})\,,
\end{aligned}
\end{equation}
with $\mathbf{R}(\mathcal{X}, \mathbf{X}_n)$ denoting the $m\times n$ matrix of $R(\mathbf{x},\mathbf{x}_i)$ for all pairs of rows between $\mathcal{X}$ and $\mathbf{X}_n$. The subscripts ``$\theta$'' and ``$\Omega$" emphasize dependencies on the corresponding parameters.  We introduce the new notation $\mathbf{R}_\theta(\mathcal{X} \mid\mathcal{D})$ to denote the conditional covariance matrix scaled by $\sigma^2$, which will be integral to our proposed DPE metric.  Fixing $g$ to a small value ensures the invertibility of $\mathbf{R}_n$, even for $\boldsymbol\theta=\mathbf{0}$.

In practice, estimated values for $\Omega$ are plugged into $\boldsymbol{\mu}_\theta(\mathcal{X}\mid\mathcal{D})$ and $\mathbf{\Sigma}_\Omega(\mathcal{X}\mid\mathcal{D})$. Estimation of $\Omega$ is driven by the GP log likelihood:
\[
    \log\mathcal{L}(\Omega \mid \mathbf{y}_n ) \propto - \frac{n}{2}\log\sigma^2 - \frac{1}{2}\log|\mathbf{R}_n| - \frac{1}{2\sigma^2}\mathbf{y}_n^\top\mathbf{R}_n^{-1}\mathbf{y}_n \ ,\ 
\]
where ``$\propto$'' indicates an additive constant has been dropped.  A closed-form MLE exists for the scale, $\hat{\sigma}^2_\theta = \frac{1}{n}\mathbf{y}_n^\top\mathbf{R}_n^{-1}\mathbf{y}_n$, providing the profile log likelihood for $\boldsymbol{\theta}$:
\begin{equation}
    \log\mathcal{L}(\boldsymbol{\theta} \mid \mathbf{y}_n, \sigma^2 = \hat{\sigma}^2_\theta) \propto - \frac{n}{2}\log\left(\mathbf{y}_n^\top\mathbf{R}_n^{-1}\mathbf{y}_n\right) - \frac{1}{2}\log|\mathbf{R}_n| \ .\ \label{eq:ploglik}
\end{equation}
The maximum likelihood estimator is $\hat{\boldsymbol{\theta}}=\text{argmax} \log \mathcal{L}(\boldsymbol{\theta} \mid \mathbf{y}_n,\sigma^2=\hat{\sigma}_\theta^2)$, which must be found numerically. Henceforth we will adopt this profile log likelihood and replace $\sigma^2$ in Eq.~(\ref{eq:gppost}) with $\hat{\sigma}^2_\theta$.

The penalized log likelihood is generally defined by
\begin{equation} \label{eq:penalized}
    Q\left(\boldsymbol\theta \mid \mathbf{y}_n, \sigma^2 = \hat{\sigma}_\theta^2\right) = \log\mathcal{L}(\boldsymbol{\theta} \mid \mathbf{y}_n, \sigma^2 = \hat{\sigma}_\theta^2) - np_{\lambda}(\boldsymbol\theta)\ ,\
\end{equation} 
for some penalty function $p_\lambda$ that is nondecreasing for $\boldsymbol\theta > 0$. The penalty function involves a tuning parameter $\lambda \geq 0$ that controls the relative importance of the penalization; larger $\lambda$ intend to produce a $\hat{\boldsymbol\theta}$ that is shrunk towards 0. 
For a specified $\lambda$, the penalized MLE (pMLE) is $\hat{\boldsymbol\theta} = \text{argmax} \;Q(\boldsymbol\theta \mid \mathbf{y}_n, \sigma^2 = \hat{\sigma}_\theta^2)$, which must be found numerically. Throughout, we use the LASSO penalty, $p_\lambda(\boldsymbol\theta) = \lambda\sum_{p=1}^d |\theta_p|$, although the approach is not restricted to this choice (more on this in Section \ref{sec:simulation}).  Our approach differs from \cite{li2005} because we employ a small fixed nugget which helps to stabilize the computations. \cite{yi2011penalized} also worked with a penalized likelihood like \eqref{eq:penalized} but they focused on stochastic simulators which requires nugget estimation. Our goal is to find a surrogate that interpolates the training data as well as possible, and estimating the nugget can actively work against that.

Tuning parameter selection plays a vital role in penalized estimation \citep{fan2013tuning, arlot2010cv}. For any such strategy, it is important to ensure the entire tuning parameter space is explored, which for us means identifying the smallest value of $\lambda$ that forces $\boldsymbol{\theta}=\mathbf{0}$. Because \cite{li2005} omit a nugget term, their model does not permit $\boldsymbol{\theta}=\mathbf{0}$. Consequently, the smallest $\lambda$ that yields $\boldsymbol{\theta}=\mathbf{0}$ is not defined in their setting. For completeness,  \Cref{app:max-lambda} derives this boundary value using the Karush-Kuhn-Tucker (KKT) optimality conditions \citep{kuhn2013nonlinear}, ensuring that the tuning parameter space is fully explored.

CV is a popular selection strategy that first partitions the data into two sets: training data $\mathcal{D}^t$ and validation data $\mathcal{D}^v$. Then penalized estimation of $\boldsymbol{\theta}$ is performed using $\mathcal{D}^t$ across a range of $\lambda$ values, each producing a prediction mean, $\boldsymbol{\mu}_\lambda (\mathbf{X}^v \mid \mathcal{D}^t)$, and prediction covariance, $\boldsymbol{\Sigma}_\lambda(\mathbf{X}^v\mid \mathcal{D}^t)$, at the validation set's input locations, $\mathbf{X}^v$.  Replacement of the subscript ``$\theta$'' with ``$\lambda$'' is meant to emphasize how the tuning parameter is the driver of the resulting pMLE values. 
The quality of these penalized estimators is evaluated on $\mathcal{D}^v$ using some metric, say $C(\mathcal{D}^v \, | \, \mathcal{D}^t, \lambda)$, involving $\boldsymbol{\mu}_\lambda (\mathbf{X}^v \mid \mathcal{D}^t)$ and/or $\boldsymbol{\Sigma}_\lambda(\mathbf{X}^v \mid \mathcal{D}^t)$. The $\lambda$ that minimizes the metric, denoted $\lambda^*$, is selected and used to estimate $\boldsymbol{\theta}$ with the full data, $\mathcal{D}$, to produce the final surrogate. A related strategy is $K$-fold CV, in which $\mathcal{D}$ is partitioned into $K$ sets, or folds, denoted by $\mathcal{D}_1, \dots, \mathcal{D}_K$, each of size $n_v=n/K$. Note, $n$-fold CV makes each $\mathcal{D}_k$ a single data point. The above CV procedure is performed for $k=1,\dots,K$, setting $\mathcal{D}^v=\mathcal{D}_k$ and $\mathcal{D}^t=\mathcal{D} \setminus \mathcal{D}_k \equiv \mathcal{D}_{-k}$, producing $K$ metrics $C(\mathcal{D}_k \mid \mathcal{D}_{-k},\lambda)$. These metrics are typically averaged at each $\lambda$ value to produce the function $C(\lambda)=K^{-1}\sum_{k=1}^K C(\mathcal{D}_k \mid \mathcal{D}_{-k},\lambda)$. The optimal $\lambda^*$ is the one that minimizes $C(\lambda)$.  

It may be desired to introduce even more penalization than that under $\lambda^*$. A popular approach is the one-standard-error rule that chooses the largest $\lambda$ whose $C(\lambda)$ is within one standard error of $C(\lambda^*)$. Applying the 1SE rule is fairly straightforward: identify $\lambda^*$, calculate $\text{SD}(\lambda^*)$, the estimated standard deviation of the $C(\mathcal{D}_k \mid \mathcal{D}_{-k},\lambda^*)$, and then calculate the estimated standard error, $\text{SE}(\lambda^*)=\text{SD}(\lambda^*)/\sqrt{K}$. The new $\lambda$ value is then
\begin{equation*}
\lambda_{1\text{SE}}= \max \left\{ \lambda \ | \ C(\lambda) \le C(\lambda^*) + \text{SE}(\lambda^*) \right\}. 
\end{equation*}
Clearly $\lambda_{\text{1SE}} \geq \lambda^*$ so the corresponding penalized estimator will experience at least as much, if not more, shrinkage compared to that under $\lambda^*$. However, as shown in Figure \ref{fig:figure2}, sometimes $\text{SE}(\lambda^*)$ can be large, which will induce excessive penalization.

The choice of the CV metric is nontrivial, especially on small data sets. In fact, an additional contribution of this paper is a discussion of why popular CV metrics can have erratic behavior for penalized MLEs of GPs. The chosen CV metric should target the statistical properties of interest and become inflated if said properties are not met. An intuitive metric is prediction error (PE):
\[
    \text{PE}(\mathcal{D}_k \mid \mathcal{D}_{-k}, \lambda) = (\mathbf{y}_k - \boldsymbol{\mu}_{\lambda,k})^\top(\mathbf{y}_k - \boldsymbol{\mu}_{\lambda,k})\ ,\
\]
where we simplify notation with $\boldsymbol{\mu}_{\lambda,k} \equiv \boldsymbol{\mu}_\lambda(\mathbf{X}_k \mid \mathcal{D}_{-k})$. This metric was recommended by \cite{li2005}, \cite{yi2011penalized}, and \cite{zhang2020penalized}, but the motivating example in Section~\ref{sec:intro} demonstrated its potential issues. In particular, PE is limited in its assessment of the predictive surrogate because it ignores UQ.  It can encourage $\hat{\boldsymbol{\theta}}$ that produce narrower prediction intervals which might fail to cover the truth.

One CV metric that incorporates UQ is Mahalanobis distance \citep[MD;][]{mahalanobis1936md},
a popular metric for Normally distributed data that accounts for uncertainty:
\[
    \begin{aligned}
    \textrm{MD}(\mathcal{D}_k \mid \mathcal{D}_{-k}, \lambda) 
    &=(\mathbf{y}_k-\boldsymbol{\mu}_{\lambda,k})^\top \boldsymbol{\Sigma}_{\lambda,k}^{-1}(\mathbf{y}_k-\boldsymbol{\mu}_{\lambda,k})\ ,\
\end{aligned}
\]
where $\boldsymbol{\Sigma}_{\lambda,k}\equiv \boldsymbol{\Sigma}_{\lambda}(\mathbf{X}_k \mid \mathcal{D}_{-k}) = \hat{\sigma}^2_{\lambda,-k}\mathbf{R}_\lambda(\mathbf{X}_k \mid \mathcal{D}_{-k})$ with $\hat{\sigma}^2_{\lambda,-k}=\left(\mathbf{y}_{-k}^\top\mathbf{R}_{\lambda,-k}^{-1}\mathbf{y}_{-k}\right)/(n-n_v)$. 
If the estimates for $\boldsymbol{\theta}$ and $\sigma^2$ equal their true values, this quantity follows a $\chi^2$ distribution with $n_v$ degrees of freedom. Another metric that incorporates UQ is Score \citep{gneiting2007score}:
\[
    \begin{aligned}
    \textrm{Score}(\mathcal{D}_k \mid \mathcal{D}_{-k}, \lambda)&=\textrm{MD}(\mathcal{D}_k \mid \mathcal{D}_{-k}, \lambda)+\log|\boldsymbol{\Sigma}_{\lambda,k}|\ ,\
\end{aligned}
\]
which is proportional to the negative log likelihood of the penalized estimates evaluated on $\mathcal{D}_k$. While closely related to Mahalanobis distance, adding $\log|\boldsymbol{\Sigma}_{\lambda,k}|$ can sometimes help to avoid surrogates with large uncertainty \citep{gramacy2020}. Choosing $n_v \geq 2$ is recommended for MD and Score to incorporate both the predicted variances and covariances in the assessment of the surrogate. In practice, smaller values of these metrics are considered to be indicative of a better fitting surrogate.

\section{Decorrelated Prediction Error}\label{sec:method}
The Forrester example in Section~\ref{sec:intro} had a profile log likelihood that was both flat and maximized for large values of $\theta$.  Large $\theta$ values can be detrimental to surrogate performance as they shrink the correlation among responses.  Specifically, for large enough $\theta$, $\mathbf{R}(\mathcal{X},\mathbf{X}_n) = \mathbf{0}$ and $\mathbf{R}_n^{-1} \approx \mathbf{I}$, leading to $\boldsymbol{\mu}_\theta(\mathcal{X} \mid \mathcal{D}) = \mathbf{0}$ and $\boldsymbol{\Sigma}_\theta(\mathcal{X} \mid \mathcal{D})\approx \hat{\sigma}^2_\theta\mathbf{I}$, where $\hat{\sigma}^2_\theta=\mathbf{y}_n^\top\mathbf{y}_n$. The exception is when a row of $\mathbf{X}_n$ is included in $\mathcal{X}$, for which $\mu_\theta(\mathbf{x}_i \mid \mathcal{D}) \approx y_i$ and $\boldsymbol{\Sigma}_\theta(\mathcal{X} \mid \mathcal{D})\approx
0$. Details of these derivations and other results in this section may be found in \Cref{app:section3}.

Penalization encourages smaller values of $\theta$, but if it goes too far it can be detrimental too. Increasing $\lambda$ shrinks $\theta$, producing smoother $\boldsymbol{\mu}_\theta(\mathcal{X} \mid \mathcal{D})$.  As $\lambda \to \infty$, $\theta \to 0$, resulting in $\mathbf{R}(\mathcal{X},\mathbf{X}_n)~\approx~\mathbf{J}$, a matrix of all ones, and $\mathbf{R}_n^{-1} \approx (g\mathbf{I}+\mathbf{J})^{-1}\approx g^{-1}\left(\mathbf{I}-\frac{1}{n+g} \mathbf{J}\right)$. The resulting surrogate has $\boldsymbol{\mu}_\theta(\mathcal{X} \mid \mathcal{D})=  \frac{n\overline{y}_{n}}{g+n}\mathbf{1}_{n_\text{test}}$, where $\overline{y}_{n}$ is the average of the training data points. The impact of increasing $\lambda$ on $\boldsymbol{\Sigma}_\theta(\mathcal{X} \mid \mathcal{D})$ is not as straightforward. As $\lambda$ initially moves away from $0$, prediction variances will start to decrease, but as $\lambda \to \infty$,   $\mathbf{\Sigma}_\theta(\mathcal{X} \mid \mathcal{D})\approx \hat{\sigma}^2_\theta\mathbf{I}$, where $\hat{\sigma}^2_\theta\approx g^{-1}\mathbf{y}_n^\top(\mathbf{I}-\frac{1}{n+g}\mathbf{J})\mathbf{y}_n$ 
becomes inflated due to our choice of $g$.  This explains the behavior shown in \Cref{fig:1f}. We will refer to surrogates with this inflated variance and $\boldsymbol{\mu}_\theta(\mathcal{X} \mid \mathcal{D})\approx \frac{n\overline{y}_{n}}{g+n}\mathbf{1}_{n_\text{test}}$ as ``nugget-dominated surrogates.''

Nugget-dominated surrogates and surrogates with excessively large $\theta$ are practically useless. Ideally, $C(\mathcal{D}_k \mid \mathcal{D}_{-k},\lambda)$ will be inflated under both surrogates and will decrease for $\lambda$ values whose penalized estimates actually improve the surrogate. Unfortunately, we have found this ideal behavior does not necessarily hold for the three metrics given in \Cref{sec:review}, particularly for small data sets where penalization has the most potential for improvement. Without a consistent and reliable tuning parameter selection method, this potential may not be realized in practice.

\begin{figure}[!ht]
    \centering
    \subfloat[$n$-fold PE\label{fig:3a}]{\includegraphics[width=0.34\linewidth, height=0.31\linewidth]{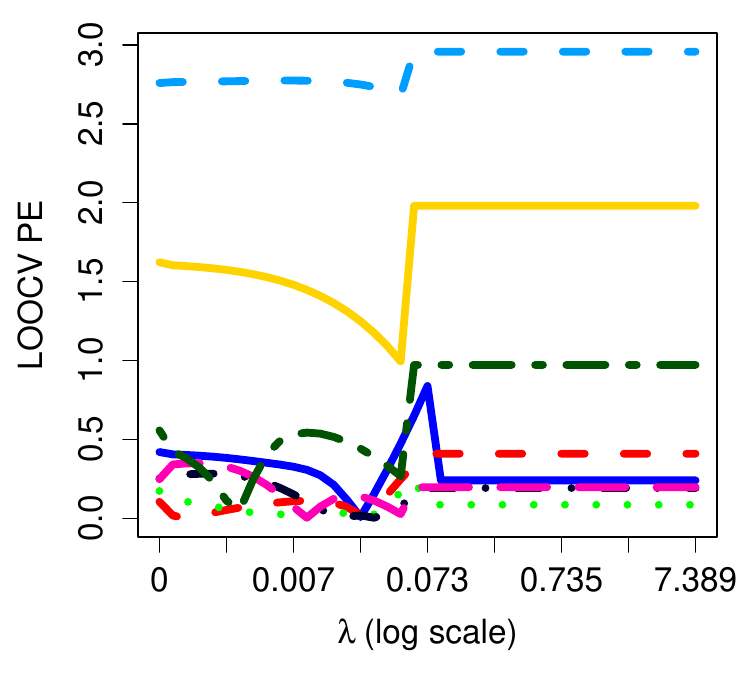}}
    \subfloat[$n$-fold Prediction at $\mathcal{D}_7$\label{fig:3b}]
{\includegraphics[width=0.34\textwidth, height=0.31\linewidth]{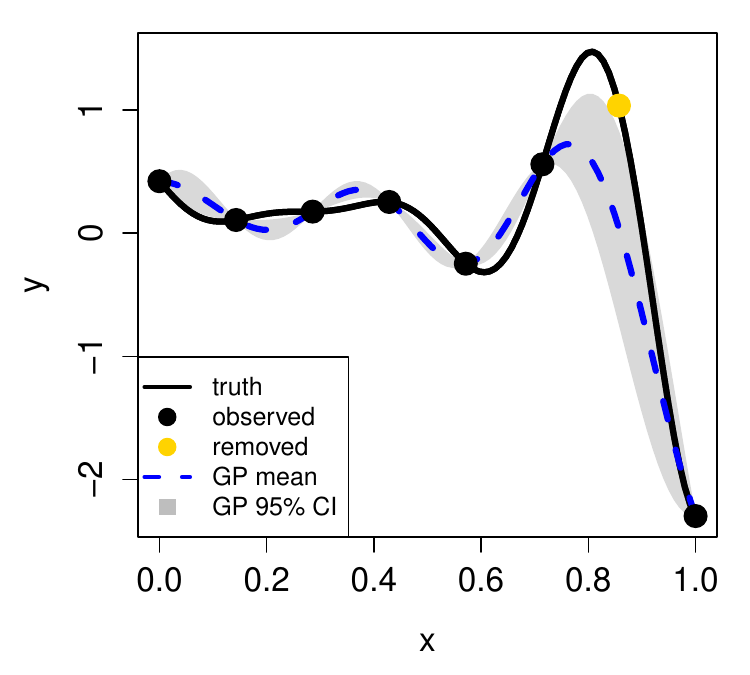}}
    \caption{(a) $n$-fold PE curves across $\lambda$ values for the Forrester function. Each colored line represents the PE curve obtained by leaving out a different data point.  (b) Surrogate predictive performance conditioned on $\mathcal{D}_{-7}$ under $\lambda=0.046$. The yellow circle marks $\mathcal{D}_7=(0.857,1.034)$, an influential point with the highest observed $y$ value.}
    \label{fig:figure3}
\end{figure}

To demonstrate these issues, we revisit the Forrester example. 
\Cref{fig:3a} shows the individual PE curves for $n$-fold CV across $\lambda$ -- one line for each leave-one-out fold (the average, $\textrm{PE}(\lambda)$, and standard errors across these lines were shown in \Cref{fig:figure2}).  There is large between-fold variability across $\lambda$. The curve with the largest values corresponds to $\mathcal{D}_{8}$, where the outlier data point ($x_8=1$) has been left out.  However, because the curve is fairly constant, it has little influence on $\textrm{PE}(\lambda)$. The second largest curve has the greatest influence on $\textrm{PE}(\lambda)$, exhibiting the most change across $\lambda$. It corresponds to $\mathcal{D}_7$, where the observation at $x_7 = 0.857$ has been left out.  This observation, highlighted by the yellow point in \Cref{fig:3b}, has the largest observed $y$ value. The surrogate based on $\mathcal{D}_{-7}$ under $\lambda=0.046$, which minimizes $\textrm{PE}(\mathcal{D}_7 \mid \mathcal{D}_{-7},\lambda)$, is shown in \Cref{fig:3b}. This ``best" surrogate was still unable to capture the observed $y_7=1.034$ and other values of the true function in a neighborhood of this observation. This serves as a reminder that there are limitations to $\mathcal{D}_{-k}$'s ability to predict at $\mathcal{D}_k$. This in turn could produce one or more folds with observations that lead to inflated $C(\mathcal{D}_k \mid \mathcal{D}_{-k},\lambda)$ values due to extrapolation.  Such folds can dominate the behavior of $C(\lambda)$, and hence the choice of $\lambda^*$. 

\begin{figure}[!ht]
    \centering
    \subfloat[PE\label{fig:4a}]{\includegraphics[width=0.34\linewidth, height=0.31\linewidth]{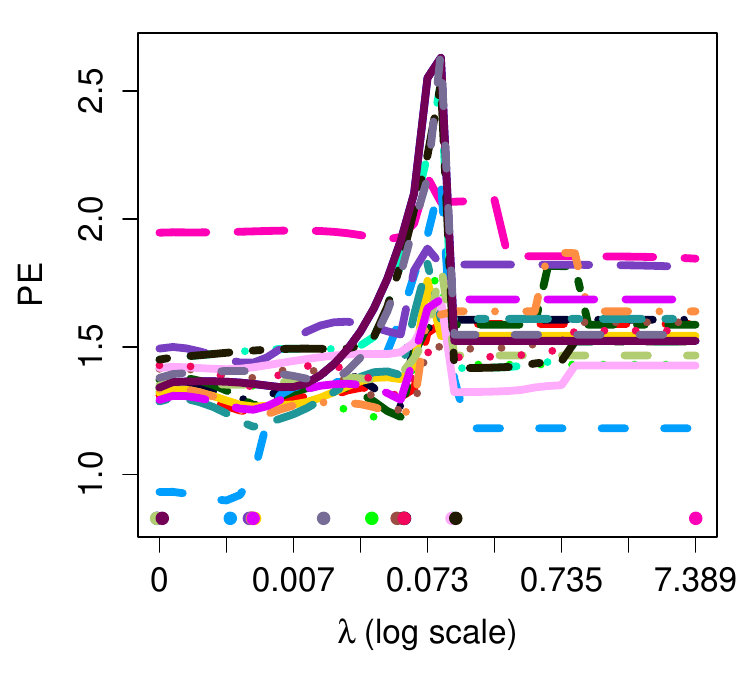}}
     \subfloat[MD\label{fig:4b}]{\includegraphics[width=0.34\linewidth, height=0.31\linewidth]{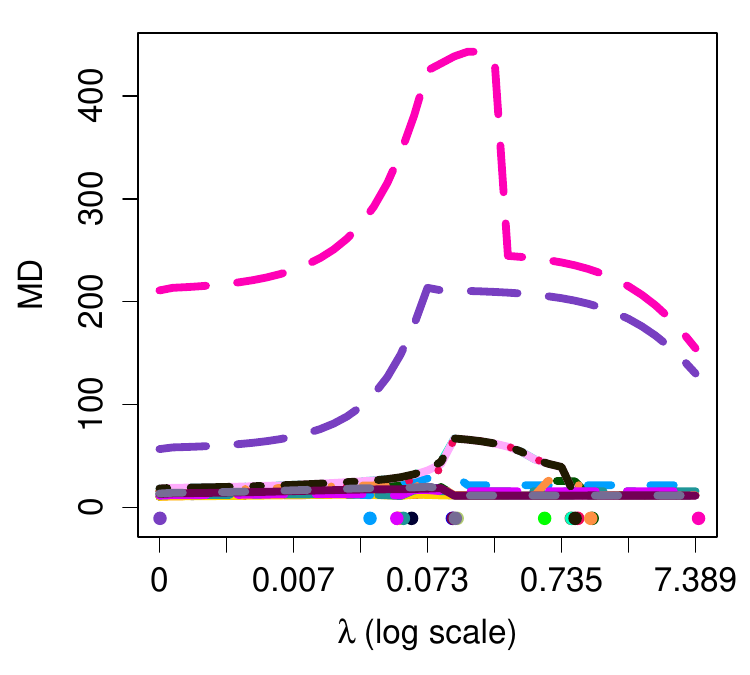}}
    \subfloat[Score\label{fig:4c}]{\includegraphics[width=0.34\linewidth, height=0.31\linewidth]{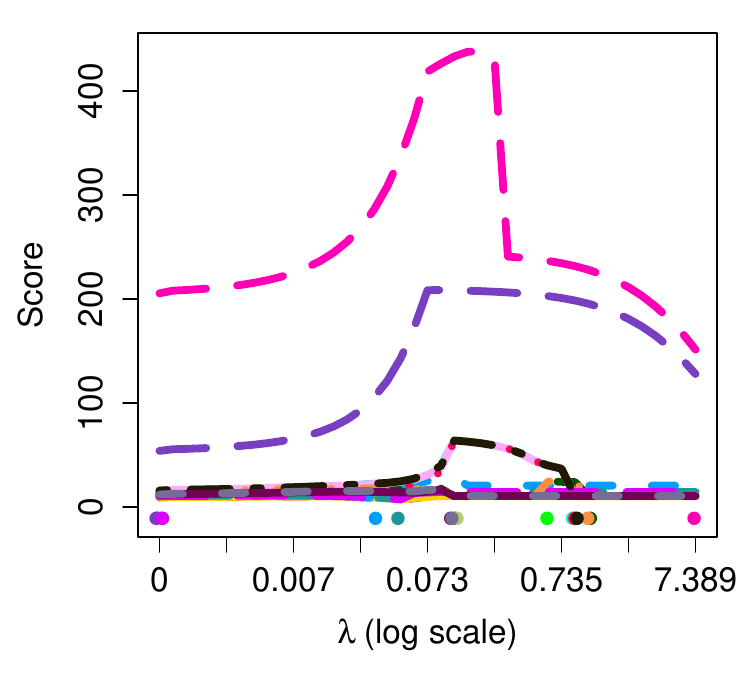}} 
    \caption{Behavior of 4-fold CV metrics across $\lambda$ values for the Forrester function.  Each line represents one of twenty repetitions with re-randomized folds.  Colored circles along the $x$-axis indicate the selected $\lambda$ value for each repetition.}
    \label{fig:forrester-kfold}
\end{figure}

Next, we repeatedly performed $4$-fold CV for the Forrester example twenty times, each with a random partitioning of the data. Each line in \Cref{fig:4a} - \ref{fig:4c} is an averaged CV metric for one of these repetitions; we considered PE, MD, and Score.  The points just above the $x$-axis are the corresponding $\lambda^*$ values. Although not shown, we also observed large between-fold variability for each of the $4$-fold CVs, similar to \Cref{fig:3a}. All three panels show large variability among the twenty $C(\lambda)$ curves and inconsistent $\lambda^*$ values across the entire range of possibilities, sometimes producing nugget-dominated surrogates.  
Across the twenty $4$-fold CVs, $\textrm{PE}(\lambda)$ sometimes chose $0 < \lambda^* \leq 0.007$, which produced surrogates comparable to \Cref{fig:1d}. However, more often it chose $\lambda^*$ that produced a surrogate that was either overly smoothed, as in \Cref{fig:1e}, or nugget-dominated, as in \Cref{fig:1f}.

Since PE uses only $\mu_\lambda(\mathbf{X}_k \mid \mathcal{D}_{-k})$, we conjectured that MD and Score, which incorporate the surrogate's UQ, could perform better.  \Cref{fig:4b} and \ref{fig:4c} show the twenty $4$-fold CV curves for MD and Score, respectively. They are nearly indistinguishable. While the distribution of $\lambda^*$ values for PE had a noticeable gap between 0.073 and 7.389, the distribution of $\lambda^*$ values for MD and Score flips the script, exhibiting a noticeable gap between $0$ and approximately $0.073$.  The large $\lambda^*$ values chosen result in nugget-dominated surrogates. The tendency for $\textrm{MD}(\lambda)$ and $\textrm{Score}(\lambda)$ to be minimized by such large $\lambda^*$ can be explained by looking at the limits of $\textrm{MD}(\mathcal{D}_k \mid \mathcal{D}_{-k}, \lambda)$ and $\textrm{Score}(\mathcal{D}_k \mid \mathcal{D}_{-k}, \lambda)$ as $\lambda \to \infty$. 

For brevity, let $\mathbf{R}_{\lambda,k}\equiv \mathbf{R}_\lambda(\mathbf{X}_k \mid \mathcal{D}_{-k})$. Using previous limiting arguments, we find 
\begin{equation*}
  \begin{aligned}
       \lim_{\lambda \to \infty} \text{MD}(\mathcal{D}_k \mid \mathcal{D}_{-k}, \lambda)&= \lim_{\lambda \to \infty}  \frac{(\mathbf{y}_k-\boldsymbol{\mu}_{\lambda,k})^\top \mathbf{R}_{\lambda.k}^{-1}(\mathbf{y}_k-\boldsymbol{\mu}_{\lambda,k})}{\hat{\sigma}^2_{\lambda,-k}}\\
    &\approx n_t \ \frac{\mathbf{y}_k^T\mathbf{y}_k+\frac{n_v^2}{n_t}\bar{y}_k^2}{\mathbf{y}_{-k}^T\mathbf{y}_{-k}-n_t\bar{y}_{-k}^2}\ .\ 
  \end{aligned}
\end{equation*} 
which no longer involves the nugget term $g$. The denominator in this approximation equals the corrected sum of squares for the responses in the $K$-th fold's training data. The numerator will become inflated when the $\mathbf{y}_k$ contains values far from $0$ and when $\bar{y}_k^2$ is large. It is also possible for MD to become deflated simply because the $\boldsymbol{y}_k$ are much closer to $0$ than $\boldsymbol{y}_{-k}$. For the Forrester data with $n_v=2$, the numerator becomes extremely unstable. Consequently, the limiting form shows that MD can be a poor metric for guarding against oversmoothing unless folds are carefully constructed or the validation sets are sufficiently large.

The limit of Score also shares similar issues:
\begin{equation*}
    \begin{aligned}
     \lim_{\lambda \to \infty}\textrm{Score}(\mathcal{D}_k \mid \mathcal{D}_{-k}, \lambda)&= \lim_{\lambda \to \infty} \textrm{MD}(\mathcal{D}_k \mid \mathcal{D}_{-k}, \lambda)+ n_v \log(\hat{\sigma}^2_{\lambda,-k}) + \log|\mathbf{R}_{\lambda,k}|\\
     &\approx  n_t \ \frac{\mathbf{y}_k^T\mathbf{y}_k+\frac{n_v^2}{n_t}\bar{y}_k^2}{\mathbf{y}_{-k}^T\mathbf{y}_{-k}-n_t\bar{y}_{-k}^2} - n_v\log(n_t) + n_v\log \left(\mathbf{y}_{-k}^T\mathbf{y}_{-k}-n_t\bar{y}_{-k}^2\right) + \log\left(1+\frac{n_v}{n_t}\right)\ .\
\end{aligned}
\end{equation*}
Again, $g$ has no effect on Score when $\lambda \to \infty$, and the additional terms provide no assurances that $\text{Score}(\lambda)$ will prevent selecting $\lambda^*$ that produces a nugget-dominated model.

When $\lambda$ becomes large enough to drive $\hat{\boldsymbol\theta}\to 0$, the variance estimate $\hat{\sigma}_\lambda^2$ inflates dramatically due to the small fixed nugget $g$.  This inflation is a useful indicator for when we are approaching a nugget-dominated surrogate. Both MD and Score, while initially appealing for their incorporation of UQ, can fail to capitalize on this information and, as a result, are prone to selecting an impractical, nugget-dominated surrogate. 

To take advantage of the inflation property of $\hat{\sigma}^2_\lambda$, we propose the decorrelated prediction error metric that incorporates both prediction accuracy and the correlation structure: 
\[
    \text{DPE}(\mathcal{D}_k \mid \mathcal{D}_{-k}, \lambda)=(\mathbf{y}_k-\boldsymbol{\mu}_{\lambda,k})^\top \mathbf{R}_{\lambda,k}^{-1}(\mathbf{y}_k-\boldsymbol{\mu}_{\lambda,k})\ .
\]
DPE gets its name from its relationship to PE, which looks at correlated responses $\mathbf{y}_k-\boldsymbol{\mu}_{\lambda,k}$. Assuming $\lambda$ produces an estimate of $\boldsymbol{\theta}$ that is close to its true value, the responses are decorrelated by pre-multiplying by $\mathbf{R}_{\lambda,k}^{-1/2}$, a square-root matrix of $\mathbf{R}_{\lambda,k}^{-1}$. DPE is also closely related to MD with the scaling factor $1/\hat{\sigma}^2_{\lambda,-k}$ removed. If $\hat{\boldsymbol{\theta}}=\boldsymbol{\theta}$, $\textrm{DPE}(\mathcal{D}_k \mid \mathcal{D}_{-k}, \lambda)$ follows a scaled $\chi^2$-squared distribution, $\sigma^2\chi^2_{n_v}$, with $n_v$ degrees of freedom. Dividing $\textrm{DPE}$ by $n_v$ would then produce an unbiased, external estimator for $\sigma^2$. Otherwise, we expect $\textrm{DPE}(\mathcal{D}_k \mid \mathcal{D}_{-k}, \lambda)$ to become inflated.  Therefore, smaller $\text{DPE}(\lambda)$ values indicate a better surrogate. This inflation property is confirmed by taking the following limit:
\begin{equation}
    \lim_{\lambda\to \infty}\text{DPE}(\mathcal{D}_k \mid \mathcal{D}_{-k}, \lambda) = \frac{1}{g} \left(\mathbf{y}_k^T\mathbf{y}_k+\frac{n_v^2}{n_t}\bar{y}_k^2\right)\ . \label{eq:dpelimit}
\end{equation}
Even for $\mathbf{y}_k$ close to zero, which would deflate MD and Score, the limit of DPE will be inflated by the $1/g$ term since $g$ is taken to be near zero. This should avoid choosing undesirable nugget-dominated surrogates.
\begin{figure}[!ht]
    \centering
     \subfloat[DPE\label{fig:5a}]{\includegraphics[width=0.34\linewidth, height=0.31\linewidth]{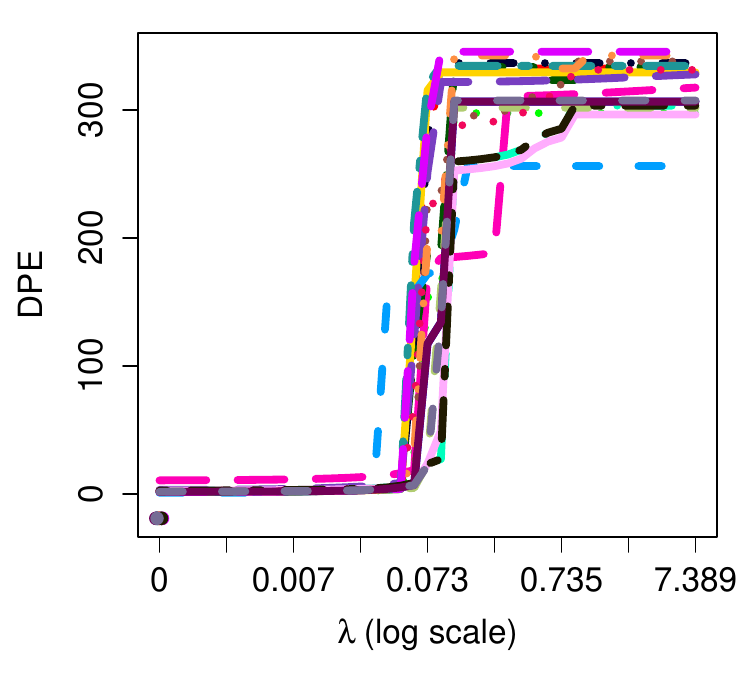}}
     \subfloat[$\lambda$ vs $\log{\hat{\theta}}$\label{fig:5b}]{\includegraphics[width=0.34\linewidth, height=0.31\linewidth]{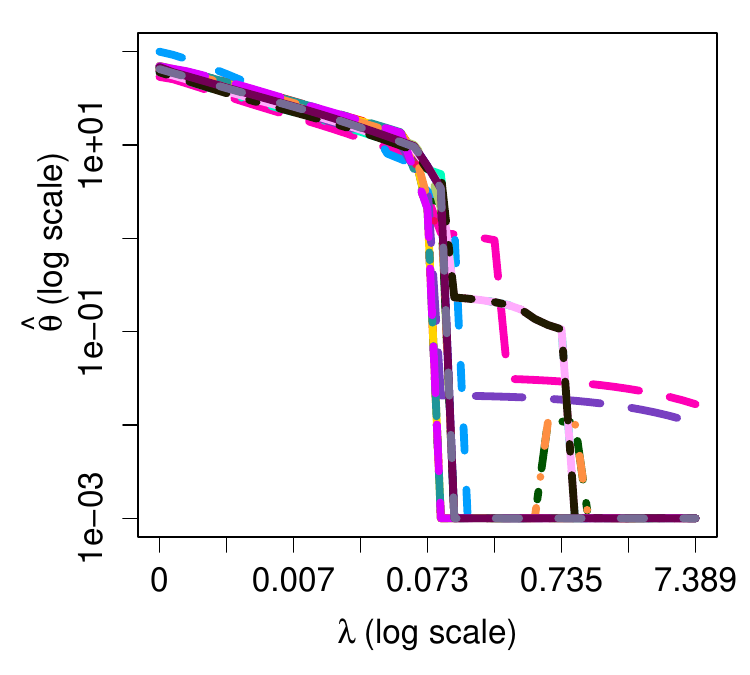}} 
     \subfloat[Selected $\lambda$\label{fig:5c}]{\includegraphics[width=0.34\linewidth, height=0.31\linewidth]{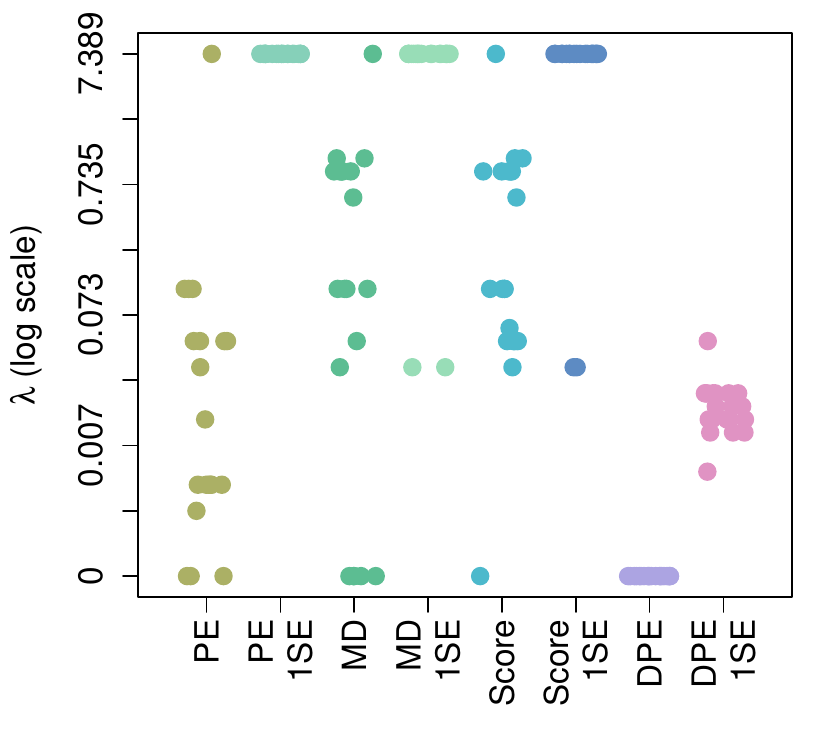}}
    \caption{(a) DPE metric  for the Forrester example across twenty 4-fold CV repetitions. (b) Resulting $\hat{\theta}$, averaged across folds.
    (c) $\lambda^\star$ and $\lambda_\textrm{1SE}$ for each of the four metrics.}
    \label{fig:forrester-kfold-dpe}
\end{figure}

Returning to the Forrester example, \Cref{fig:5a} shows the DPE metric across the same 20 repeated $4$-fold CVs from \Cref{fig:forrester-kfold}. Compared to PE, MD, and Score, the DPE curves have consistently higher values for larger $\lambda$ and consistently lower values when $\lambda<0.073$. This indicates that the DPE metric may be less sensitive to fold-to-fold variability, reducing concerns about instability in tuning parameter selection.
\Cref{fig:5b} shows the resulting $\hat{\theta}$ values, estimated for each fold and averaged across folds.  When $\lambda>0.073$, the average $\hat{\theta}$'s drop sharply toward zero, and the DPE curves in \Cref{fig:7a} become inflated, matching the expectations set by \eqref{eq:dpelimit}. DPE chose $\lambda^*=0$ for all of the $4$-fold CVs.  Moreover, the reduced fold-to-fold variability and the rapid inflation of DPE for large $\lambda$ suggests potential value in applying the 1SE rule. The $\lambda^*$ and $\lambda_{\text{1SE}}$ values selected by all four metrics are shown in \Cref{fig:5c}. Due to the high between-fold variability of PE, MD, and Score, applying the 1SE rule almost always resulted in $\lambda_{\textrm{1SE}}=7.389$. The $\lambda_{\textrm{1SE}}$ values under DPE   show remarkable consistency while avoiding overly large $\lambda$.

\section{Numerical Studies}\label{sec:simulation}

In this section, we examine the performance of tuning parameter selection with $K$-fold CV  under PE, MD, Score, and DPE on a variety of benchmark functions. Given our observations in \Cref{sec:method}, the 1SE method was only applied to the DPE metric. We also include the MLE approach with an estimated nugget, as advocated by \cite{gramacy2012nugget}.

\paragraph{Implementation details.} As the lengthscale parameter $\boldsymbol{\theta}$ lacks a closed-form solution, numerical optimization of Eq.~(\ref{eq:penalized}) is required. We use the \texttt{optim} function in R with the L-BFGS-B algorithm \citep{byrd1995}, which supports gradient-based updates and bound constraints.  We restrict the search space of $\boldsymbol{\theta}$ to the interval $[0.001, 1000]^d$. Since gradient-based optimization methods are sensitive to initial values, we use 10 random multi-starts to reduce the risk of convergence to local optima \citep{marti2003}.  Our software \citep[{\tt GPpenalty};][]{mutoh2025package} offers both the LASSO penalty introduced in \Cref{sec:review} and the smoothly clipped absolute deviation (SCAD) penalty \citep{fan2001scad}. After examining both LASSO and SCAD, we found no significant difference in performance and therefore adopted the simpler LASSO penalty for the following exercises.

\paragraph{Test functions.} We consider four different test functions that are commonly found in the GP literature: Lim Nonpolynomial \citep{limfunction}, Franke \citep{frankefunction}, Piston Simulation \citep{pistonfunction1}, and Borehole \citep{borehole}. Training and testing data were generated with random Latin hypercube sampling using the {\tt lhs} R package \citep{carnell2025lhs}. We varied data sizes based on input dimension (see \Cref{tab:function}), aiming to replicate real-world small data scenarios.  We performed $5$-fold CV for all data sets. The final surrogate for each method was estimated using the full training dataset $\mathcal{D}$ with either $\lambda^*$ or $\lambda_{\textrm{1SE}}$, then evaluated on the test data set.  We repeated this process 100 times with new Latin hypercube samples.

\begin{table}[ht]
    \centering
    \begin{tabular}{|c|c|c|c|}
    \hline
      Function & $d$ & $n$ & $n_{\text{test}}$ \\
      \hline
      Lim Nonpolynomial & 2 & 10  & 200\\
      Franke & 2 & 10 & 200\\
      Piston Simulation & 7 & 15 & 700\\
      Borehole & 8 & 15  & 800 \\
      \hline
    \end{tabular}
    \caption{Simulation settings for each test function.}
    \label{tab:function}
\end{table}

\paragraph{Validation metrics.} To assess the surrogate's prediction accuracy and UQ, we evaluated root mean squared error (RMSE) and continuous ranked probability score \citep[CRPS;][]{gneiting2007score} on the test data. RMSE is defined as
\begin{equation*}
    \textrm{RMSE}=\sqrt{\frac{1}{n_{\textrm{test}}}\sum^{n_{\textrm{test}}}_{i=1}(y_{\textrm{test},i} - \mu_{\lambda^*,i})^2}\ ,\
\end{equation*}
where $\mu_{\lambda^*,i} \equiv \mu_{\lambda^*}(\mathbf{x}_{\textrm{test},i} \mid \mathcal{D})$. 
When the predictive distribution is Gaussian with mean $\mu_{\lambda^*,i}$ and variance $\tau_{\lambda^*,i}^2=\boldsymbol{\Sigma}_{\lambda^*}(\mathbf{x}_{\textrm{test},i} \mid \mathcal{D})$,
CRPS is defined as 
\begin{equation*}
    \textrm{CRPS}=-
\frac{1}{n_\textrm{test}}\sum_{i=1}^{n_\textrm{test}} \tau_{\lambda^*,i} \left(\frac{1}{\sqrt{\pi}} - 2\phi(z_{\lambda^*,i}) - z_{\lambda^*,i} (2\Phi(z_{\lambda^*,i}) - 1) \right) \quad \textrm{for} \quad z_{\lambda^*,i} = \frac{y_{\textrm{test},i} - \mu_{\lambda^*,i}}{\tau_{\lambda^*,i}}\ ,\
\end{equation*}
where $\phi$ and $\Phi$ represent the $\mathcal{N}(0,1)$ probability density function and cumulative distribution function, respectively.  Here, we have negated the standard form from \citet{gneiting2007score}, so a lower CRPS indicates a better distributional prediction.  To demonstrate the potential for penalized MLE to improve the surrogate, we included two additional competitors: the GP with unpenalized MLE and the GP with optimal penalization, denoted pMLE*, which uses the $\lambda$ that minimized RMSE on the test data. 

\begin{figure}[!ht]
    \centering
    \textbf{Relative RMSE}
    \subfloat[Lim Nonpolynomial\label{fig:6a}]{\includegraphics[width=0.26\linewidth, height=0.29\linewidth]{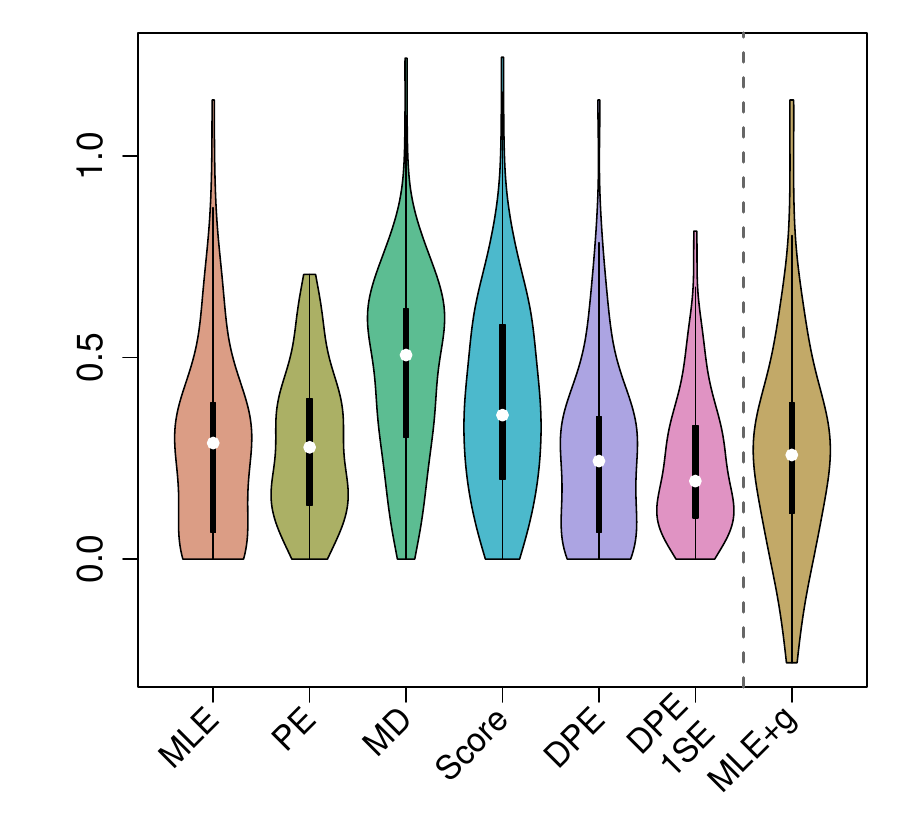}}
    \subfloat[Franke \label{fig:6b}]{\includegraphics[width=0.26\linewidth, height=0.29\linewidth]{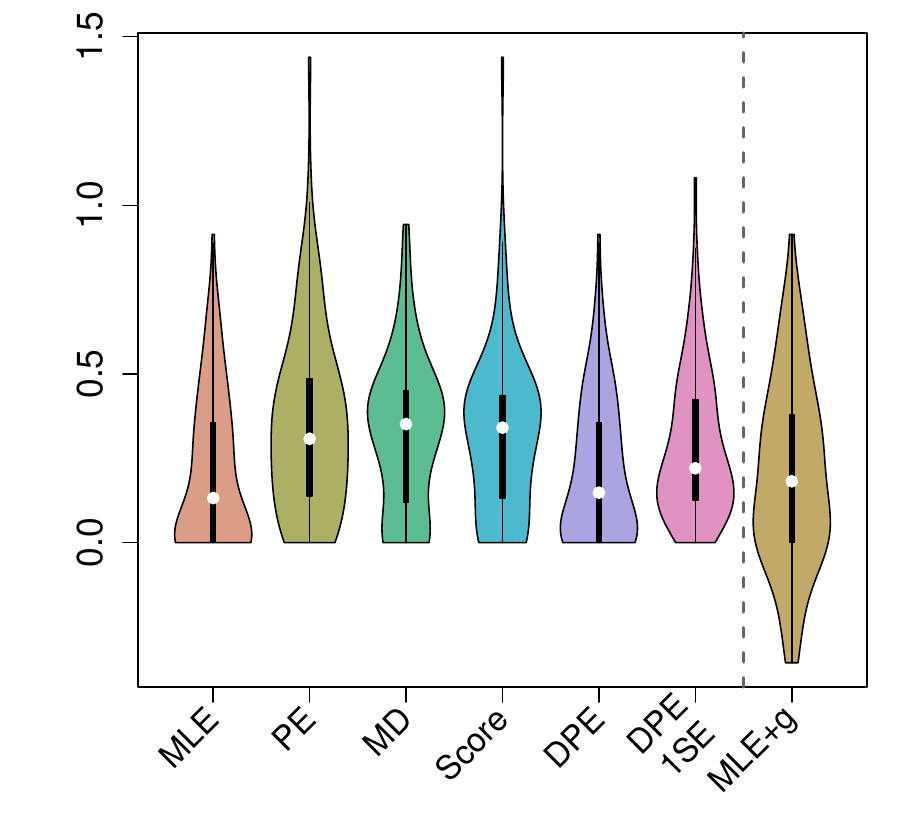}}
    \subfloat[Piston Simulation\label{fig:6c}]{\includegraphics[width=0.26\linewidth, height=0.29\linewidth]{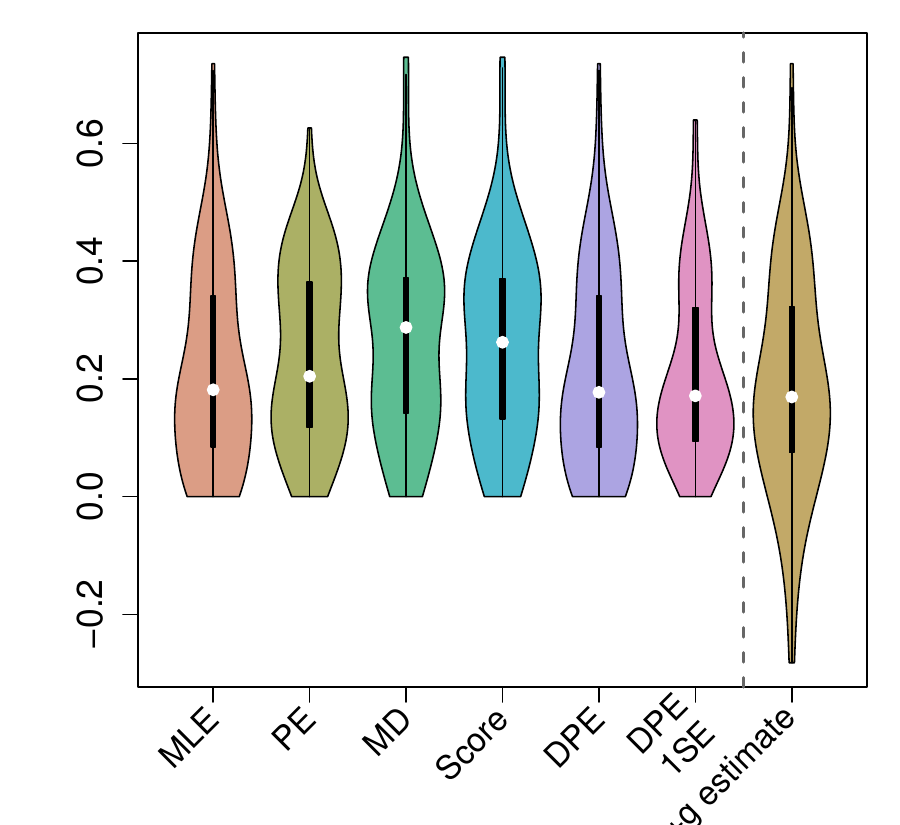}}
    \subfloat[Borehole\label{fig:6d}]{\includegraphics[width=0.26\linewidth, height=0.29\linewidth]{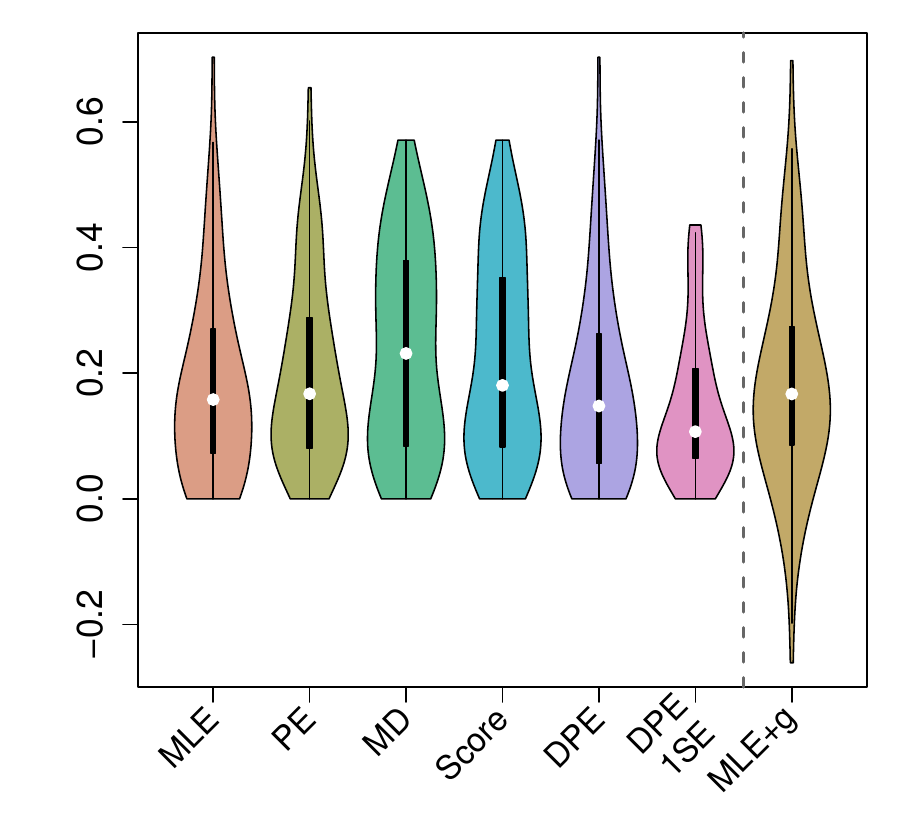}} \\
    \bigbreak
    \textbf{Relative CRPS}
    \subfloat[Lim Nonpolynomial\label{fig:6e}]{\includegraphics[width=0.26\linewidth, height=0.29\linewidth]{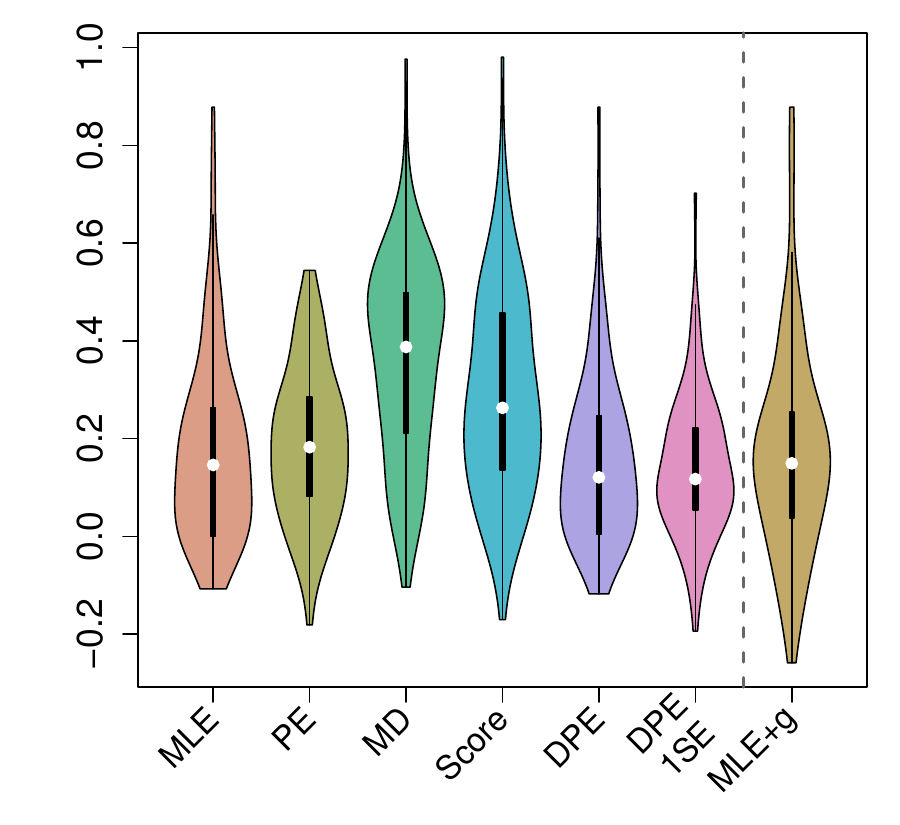}}
    \subfloat[Franke\label{fig:6f}]{\includegraphics[width=0.26\linewidth, height=0.29\linewidth]{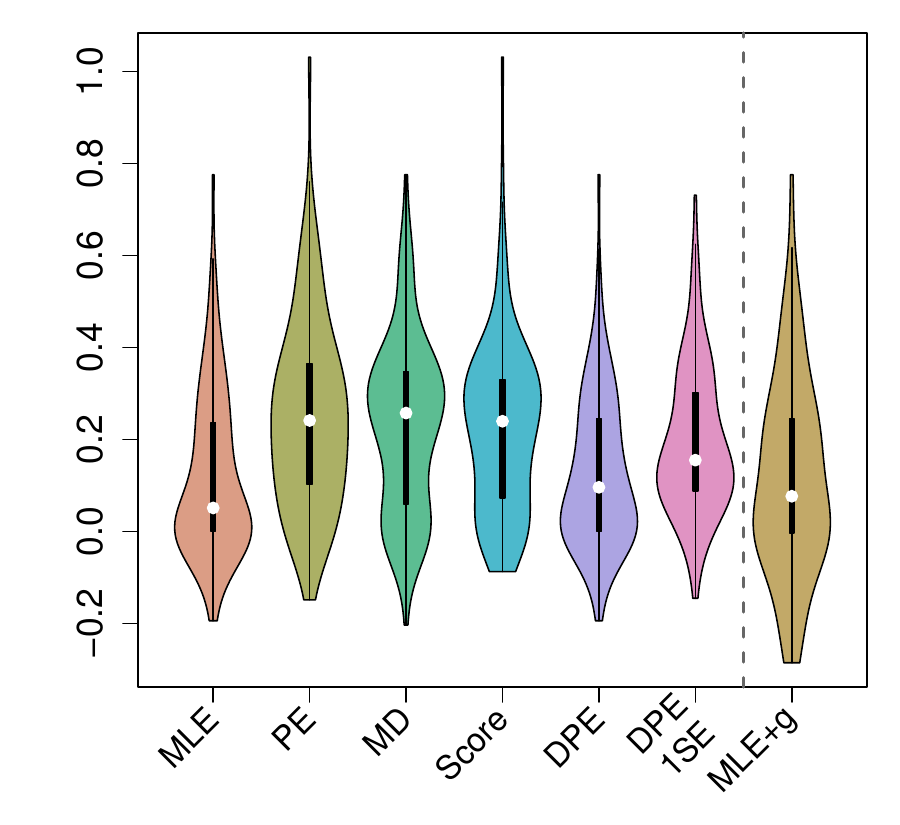}}
    \subfloat[Piston Simulation\label{fig:6g}]{\includegraphics[width=0.26\linewidth, height=0.29\linewidth]{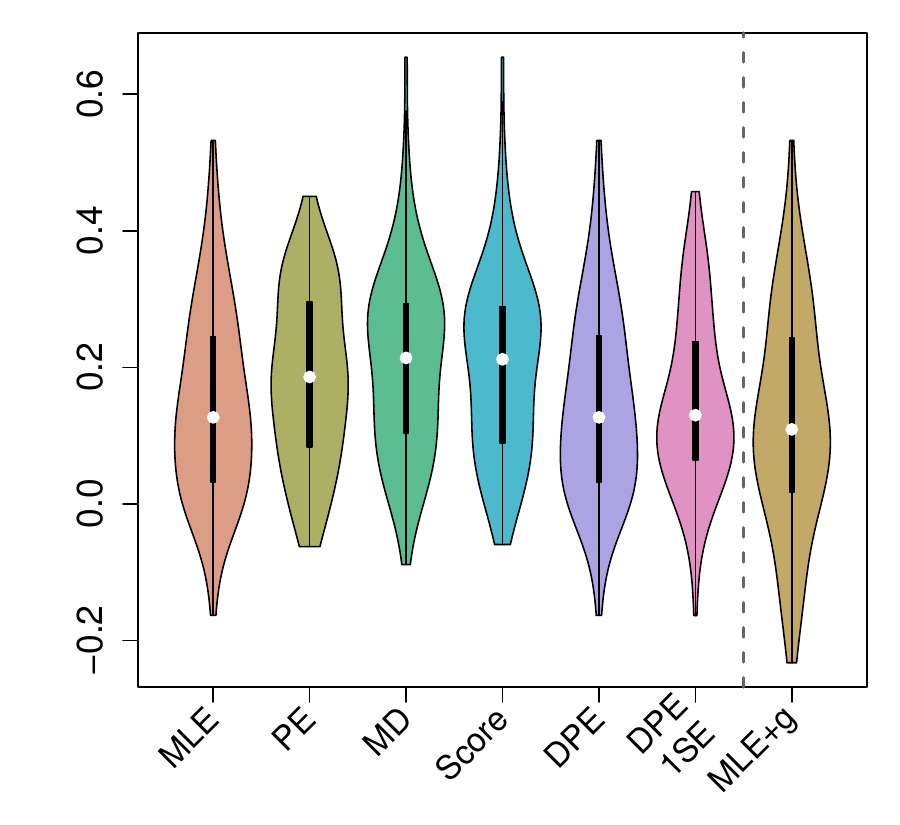}}
    \subfloat[Borehole\label{fig:6h}]{\includegraphics[width=0.26\linewidth, height=0.29\linewidth]{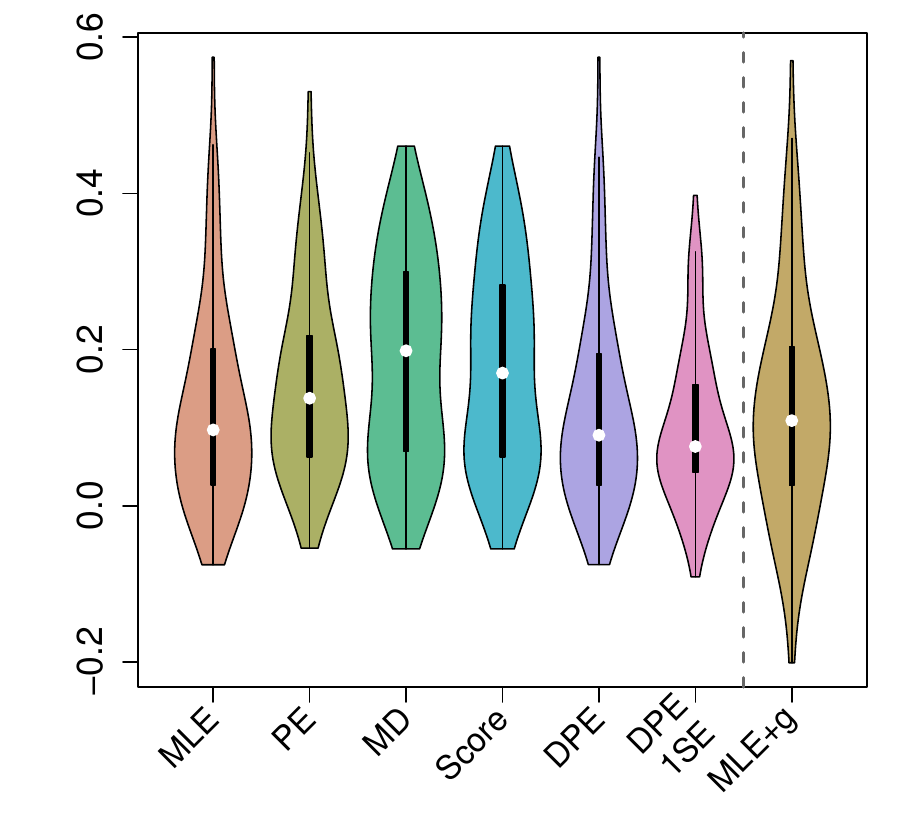}} 
     \caption{RMSE (panels (a)-(d), upper) and CRPS (panels (e)-(h), lower) evaluated across 100 repetitions. The $y$-axis shows the relative difference from pMLE*, plotted on a signed square root scale, $\text{sign}(x)\sqrt{|x|} $.}
    \label{fig:simulation}
\end{figure}

\paragraph{Results.} \Cref{fig:simulation} shows the RMSE and CRPS for MLE with both fixed $g$ and estimated $g$, and the different tuning parameter selection strategies relative to those under the optimal pMLE*. Across the board, the performances of DPE and MLE were closely aligned. MD and Score performed the worst due to their tendency to select large $\lambda$ values. PE showed slight improvement over MLE for the Lim Nonpolynomial but did similar or worse for the other functions. For all functions other than Franke, DPE 1SE provided modest, consistent improvements over MLE.  The additional penalization chosen by DPE 1SE appears undesirable for the Franke function, slightly elevating both RMSE and CRPS. For this function, MLE appears to already do quite well, so penalization seems ill-advised. Notably, DPE and DPE 1SE outperformed PE, MD, and Score metrics for all functions, especially in terms of CRPS. This performance improvement is due to DPE's tendency to select smaller values of $\lambda$. The MLE with an estimated $g$ revealed highly variable performance, sometimes improving over pMLE* but other times doing much worse than DPE 1SE. Although none of the methods were able to consistently match the performance of pMLE*, the DPE and DPE 1SE approach proved to be the most reliable.

\section{Piston Slap Noise Data} \label{sec:piston}
In this section, we analyze the piston slap noise data introduced by \cite{hoffman2003piston} and considered in \cite{li2005}. The response variable is piston slap noise, and the 6 input settings are: cylinder liner ($x_1$), location of peak pressure ($x_2$), skirt length ($x_3$), skirt profile ($x_4$), skirt ovality ($x_5$), and pin offset ($x_6$). Each run of the computer experiment required 24 hours of computation.  We copy the training/testing split from \citet{li2005} with $n=12$ and $n_\text{test} = 100$ (training data details are provided in \Cref{app:piston}). \cite{li2005} originally analyzed the data without scaling the input settings and instead set $\theta_p=\theta/\sigma_p$ where $\sigma_p$ is the sample standard deviation of the $p$-th input setting. As we do advocate for scaling the inputs, this is equivalent to us fitting an isotropic version of \eqref{eqn:cov_func} with $\theta_p=\theta$. Our analysis based on an anisotropic correlation function may be found in \Cref{app:anisotropic}.

\begin{figure}[!ht]
    \centering
     \subfloat[RMSE\label{fig:7a}]{\includegraphics[width=0.38\linewidth]{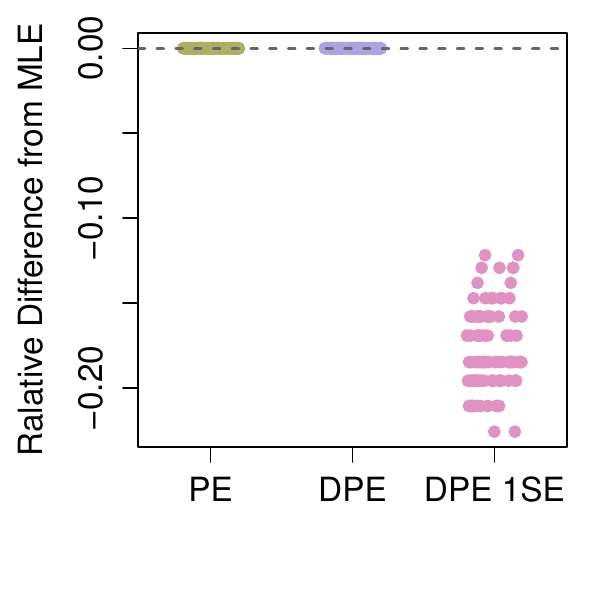}}
     \subfloat[CRPS\label{fig:7b}]{\includegraphics[width=0.38\linewidth]{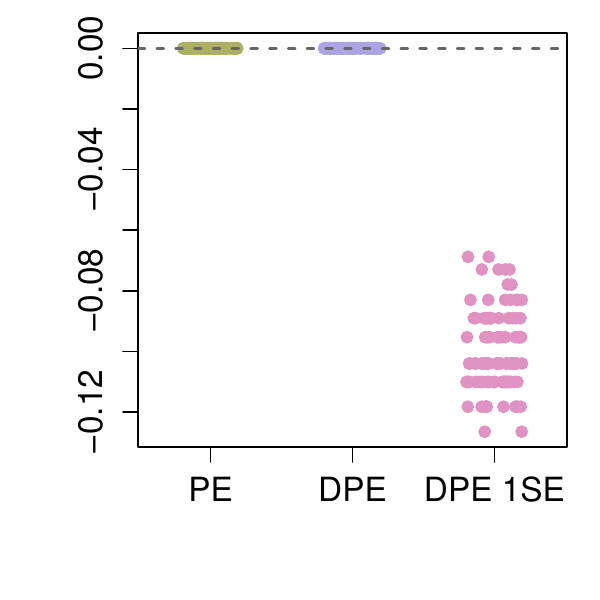}}

    \caption{Relative differences in RMSE and CRPS compared with the MLE baseline. The gray dashed line represents zero difference, corresponding to selecting $\lambda=0$.}
    \label{fig:piston}
\end{figure}

The MLE analysis on the training data resulted in an RMSE of 0.879 and a CRPS of 0.505 on the test data. Our implementation of $n$-fold CV with PE chose $\lambda=0$, hence producing the same RMSE and CRPS values as MLE. We then implemented $4$-fold CV on the $n=12$ training data with PE, DPE, and DPE 1SE. Since the performance of $4$-fold CV can depend on the construction of the folds, we repeated the procedure for 100 randomly generated folds. We evaluated performance by comparing RMSE/CRPS against the ``benchmark'' MLE values. \Cref{fig:piston} shows the relative differences in RMSE and CRPS for all 100 repetitions. A value of zero means the method matched the MLE performance and negative values indicate improvement. Both PE and DPE selected $\lambda=0$ and so matched the MLE.  The surrogate chosen by DPE 1SE, however, significantly improved over MLE for all 100 randomly generated folds. While these results highlight the potential benefits of penalization in limited sample settings, we also see a high variability in that improvement depending on the fold configuration. A more structured fold design may yield more reliable results than random partitioning.

To provide more context for this improved performance, \Cref{tab:piston_estimates} shows $\hat{\sigma}_\theta^2$ and $\hat{\theta}$ for the MLE surrogate alongside the estimates obtained at $\lambda=0.116$, which produced the greatest improvement in RMSE and CRPS among the 4-fold CV runs with DPE 1SE. Even this modest level of penalization leads to distinctly different estimates, with $\hat{\theta}$ shrunk significantly from $18.992$ to $1.453$. This underscores how small amounts of penalization can reshape hyperparameter estimates.  

\begin{table}[H]
    \centering
    \begin{tabular}{|c||c|c|c|}
                \hline
                    & MLE &  Best $4$-fold CV   \\
                    \hline
                 $\lambda$  & 0& 0.116   \\
                 \hline
                 $\hat{\sigma}^2_\theta$ & 0.917  & 1.177 \\
                 \hline
                 $\hat{\theta}$ & 18.992 & 1.453   \\
                 \hline 
                \end{tabular}
    \caption{Parameter estimates from the piston slap noise data for MLE and a $4$-fold CV providing improved RMSE/CRPS (one of the pink dots in \Cref{fig:piston}).
    }
    \label{tab:piston_estimates}
\end{table}

\section{Discussion} \label{sec:discussion}
This article examined the penalized likelihood framework for GPs, originally proposed by \citet{li2005}, discussing common pitfalls and proposing potential remedies.  We demonstrated the importance of including a nugget for numerical stability, but then showed how it may have unintended consequences on other parameter estimates, leading to ``nugget-dominated surrogates.''
Popular metrics for evaluating GPs, including PE, MD, and Score, often struggle to consistently identify reliable tuning parameters under $K$-fold CV with small data sets. In particular, PE's ignorance of UQ tended to produce surrogates whose UQ failed to capture the true function. MD and Score, which incorporate UQ, were shown to have issues caused by inclusion of the nugget effect, leading to their tendency to choose nugget-dominated surrogates. We ultimately proposed a new metric, DPE, that accounts for UQ and avoids $\lambda$ that produce nugget-dominated surrogates. In our numerical studies across multiple test functions, we found DPE to be conservative in the amount of recommended penalization; combining it with the 1SE rule often produced smoother predictive surfaces with improved UQ compared to MLE.

The potential for penalized estimation of GPs to improve upon MLE is greatest with small training data, such as with resource-heavy computer experiments. However, this potential can only be realized with a consistent and reliable tuning parameter selection strategy. CV remains the most widely used approach for tuning parameter selection, but it relies heavily on the choice of metric and construction of the folds. We have given analytical and numerical arguments for why DPE and DPE 1SE are preferable CV metrics, but selection of the best $\lambda$ from limited observations remains a difficult problem.

Based on our observations on the Forrester function (see \Cref{fig:figure3}) and piston slap noise data, we find that fold construction may have a greater impact on CV performance than the choice of metric, particularly for small data settings. The quality of the folds also depends on the structure of the training data. For example, the 12 training observations of the piston slap noise data (Appendix \ref{app:piston}) have poor one-dimensional projection properties. Variables 4 and 5 only have three unique values, and the remaining variables only have six unique observed values. Any partitioning of this data is likely to create CV folds with even weaker projection properties. The most important direction for future work in this area is developing more effective design construction strategies for both the training data and its CV folds.

\bibliographystyle{apalike}
\bibliography{ref}

\newpage
\begin{center}
    {\Large\bf Appendix}
\end{center}
\setcounter{section}{0}
\setcounter{equation}{0}
\setcounter{figure}{0}

\appendix
\section{Maximum tuning parameter values} \label{app:max-lambda}
When selecting a tuning parameter, it is common to evaluate a grid of values to identify the optimum. To ensure the entire parameter space is explored, we must determine an appropriate maximum value. However, setting this upper bound is often challenging without a principled rule of thumb, as it may depend on the data or other factors. To address this, we utilize the Karush-Kuhn-Tucker (KKT) optimality conditions \citep{kuhn2013nonlinear}. This approach is widely adopted in linear regression with a LASSO penalty, including implementations such as the R package {\tt glmnet} \citep{friedman2010regularization}; see also
\cite{hastie2015statistical}. 

In our framework, the threshold is derived from the gradient of the penalized log-likelihood function defined in \Cref{eq:penalized} evaluated at $\lambda=0$, which corresponds to \Cref{eq:ploglik}. Since the objective function $\mathcal{L}(\theta)$ is non-differentiable at $\theta_p=0$, the maximum tuning parameter is defined as
\[
\lambda_{\max} = \frac{1}{n}\max_p \left|\frac{\partial \log\mathcal{L}(\theta_p)}{\partial \theta_p}(0)\right|,
\]
where \[\frac{\partial\log \mathcal{L}(\theta_p)}{\partial \theta_p} = -\frac{1}{2}\text{tr}\left(\mathbf{R}_n^{-1}\frac{\partial \mathbf{R}_n}{\partial \theta_p}\right) + \frac{n}{2\mathbf{y}_n^\top\mathbf{R}_n^{-1}\mathbf{y}}\mathbf{y}_n^\top \mathbf{R}_n^{-1}\frac{\partial \mathbf{R}_n}{\partial \theta_p}\mathbf{R}_n^{-1}\mathbf{y}_n,\]
 and   
\[\frac{\partial R_{ij}}{\partial \theta_p}=-(x_{ip}-x_{jp})^2\exp{\left(-\sum^d_{k=1}\theta_k(x_{ik}-x_{jk})^2\right)} .\]

\section{Detailed Derivations for Section 3} \label{app:section3}
\subsection{Inverse covariance matrix under extreme parameter values}\label{sec:Appendix_Inverse}
We consider an isotropic case, where the kernel function is defined as $\mathbf{R}_n^{ij} = R(x_i, x_j) = \mathrm{exp}\left(-\theta (x_i - x_j)^2\right)$. When the penalized estimate for $\theta$ approaches 0, $R(x_i, x_j) \approx 1$, which implies $\mathbf{R}(\mathcal{X}, \mathbf{X}_n)\approx \mathbf{J}$. Consider now the modified covariance function, $R(x_i,x_j) = \mathrm{exp}\left(-\theta (x_i - x_j)^2\right) + g\mathbb{I}_{(i=j)}$. With the inclusion of the $g$ term, as $\theta\to 0$, the covariance function becomes 
\[\mathbf{R}_n\approx g\mathbf{I}+\mathbf{J}=g\mathbf{I} + \mathbf{1}\mathbf{1}^\top. \] 
Applying the Sherman-Morrison formula, we obtain
\begin{equation*}
    \begin{aligned}
        \mathbf{R}^{-1}_n&=(g\mathbf{I} + \mathbf{1}\mathbf{1}^\top)^{-1} \\
    &=\frac{1}{g}\left(\mathbf{I}-\frac{1}{g+n}\mathbf{J} \right) \\
    \end{aligned}
\end{equation*}

\subsection{Limiting form of the predictive mean function}
The predictive mean function under $K$-fold CV is defined as 
\[
\boldsymbol{\mu}_{\lambda, k} = \mathbf{R}(\mathbf{X}^v, \mathbf{X}^t)\mathbf{R}_{n_t}^{-1} \mathbf{y}_{-k}
\]
As $\lambda\to \infty$,
\[
\begin{aligned}
    \lim_{\lambda \to \infty}\boldsymbol{\mu}_{\lambda, k} &= \mathbf{J}_{n_v\times n_t} \left(\frac{1}{g}\left(\mathbf{I}_{n_t}-\frac{1}{g+n_t}\mathbf{J}_{n_t}\right)\right) \mathbf{y}_{-k} \\
    &= \frac{n_t\overline{y}_{-k}}{g+n_t}\mathbf{1}_{n_v}
\end{aligned}
\]
where $\overline{y}_{-k}$ is the average of the training data points. 

\subsection{Limiting form of the predictive covariance matrix}
The predictive covariance matrix under $K$-fold CV is given by
\begin{equation*}
    \mathbf{R}_\lambda(\mathbf{X}_k \mid \mathcal{D}_{-k})=\mathbf{R}(\mathbf{X}^v,\mathbf{X}^v)- \mathbf{R}(\mathbf{X}^v, \mathbf{X}^t)\mathbf{R}_{n_t}^{-1} \mathbf{R}(\mathbf{X}^t,\mathbf{X}^v)
\end{equation*}
Let $\mathbf{R}_{\lambda,k}\equiv \mathbf{R}_\lambda(\mathbf{X}_k \mid \mathcal{D}_{-k})$. As $\lambda\to \infty$,
 \begin{equation*}
        \begin{aligned}
        \lim_{\lambda \to \infty}\mathbf{R}_{\lambda,k}&=\mathbf{J}_{n_v} + g\mathbf{I}_{n_v}-\mathbf{J}_{n_t\times n_v}\left(\frac{1}{g}\left(1-\frac{1}{g+n_t}\mathbf{J}_{n_t}\right)\right)\mathbf{J}_{n_v\times n_t} \\
		&=g\left(\frac{1}{g+n_t}\mathbf{J}_{n_v} + \mathbf{I}_{n_v}\right) \ . \
        \end{aligned}
    \end{equation*}

\subsection{Limiting form of Mahalanobis distance}
Mahalobis distance (MD) is defined as 
\begin{equation*}
    \begin{aligned}
        \textrm{MD}(\mathcal{D}_k \mid \mathcal{D}_{-k}, \lambda) &=(\mathbf{y}_k-\boldsymbol{\mu}_{\lambda,k})^\top \boldsymbol{\Sigma}_{\lambda,k}^{-1}(\mathbf{y}_k-\boldsymbol{\mu}_{\lambda,k})\\
        &=\frac{1}{\hat{\sigma}^2_{\lambda,-k}}(\mathbf{y}_k-\boldsymbol{\mu}_{\lambda,k})^\top \mathbf{R}^{-1}_{\lambda, k}(\mathbf{y}_k-\boldsymbol{\mu}_{\lambda,k}) \ ,\
    \end{aligned}
\end{equation*}
where $\hat{\sigma}^2_{\lambda,-k}=\mathbf{}{y}_{-k}^\top\mathbf{R}_{-k}^{-1}\mathbf{y}_{-k}/n_t$. The limit of the denominator is 
\[\begin{aligned}
     \lim_{\lambda \to \infty} \hat{\sigma}^2_{\lambda,-k} &= \frac{1}{n_t}\lim_{\theta \to 0} \mathbf{y}_{-k}^T\mathbf{R}_{n_t}^{-1}\mathbf{y}_{-k}\\
    &=\frac{1}{n_t}\mathbf{y}_{-k}^T\left(\mathbf{J}_{n_t}+g\mathbf{I}_{n_t}\right)^{-1}\mathbf{y}_{-k}\\
    &=\frac{1}{g} \ \frac{1}{n_t}\mathbf{y}_{-k}^T\left(\mathbf{I}_{n_t}-\frac{1}{n_t+g}\mathbf{J}_{n_t}\right)\mathbf{y}_{-k}\ .\
\end{aligned}\] 
This limit is always positive since $\left(\mathbf{I}_{n_t}-\frac{1}{n_t+g}\mathbf{J}_{n_t}\right)$ is positive definite. Therefore, $\lim_{\lambda \to \infty} \text{MD}(\mathcal{D}_k \mid \mathcal{D}_{-k}, \lambda)$ can be found by taking the ratio of limits for the numerator and denominator.

As $\lambda \to \infty$, $\mathbf{R}_{\lambda,k} =g\left(\frac{1}{g+n_t}\mathbf{J}_{n_v} + \mathbf{I}_{n_v}\right)$. Applying the Sherman-Morrison formula, we obtain
    \begin{equation*}
        \begin{aligned}
            \mathbf{R}^{-1}_{\lambda,k} 
			&=\frac{1}{g} \left(\mathbf{I}_{n_v}-\frac{1}{g+n}\mathbf{J}_{n_v}\right) \ .\ 
        \end{aligned}
    \end{equation*}
    By the continuity of matrix multiplication
\begin{align} \label{eq:limit-numerator}
    \lim_{\lambda \to \infty} (\mathbf{y}_k-\boldsymbol{\mu}_{\lambda,k})^\top \mathbf{R}_{\lambda.k}^{-1}(\mathbf{y}_k-\boldsymbol{\mu}_{\lambda,k})=\frac{1}{g}\left(\mathbf{y}_k - \frac{n_t\bar{y}_{-k}}{n_t+g}\mathbf{1}_{n_v}\right)^T\left(\mathbf{I}_{n_v}-\frac{1}{n+g}\mathbf{J}_{n_v}\right)\left(\mathbf{y}_k - \frac{n_t\bar{y}_{-k}}{n_t+g}\mathbf{1}_{n_v}\right)\ .\
\end{align} 
By the fact that
\[\mathbf{I}=\left(\mathbf{I}_{n_v}+\frac{1}{n_t+g}\mathbf{J}_{n_v}\right)\left(\mathbf{I}_{n_v}-\frac{1}{n+g}\mathbf{J}_{n_v}\right) \ . \ \]
This holds since $n=n_t+n_v$. Inserting this quantity to \Cref{eq:limit-numerator} yields,
\[\lim_{\lambda \to \infty} (\mathbf{y}_k-\boldsymbol{\mu}_{\lambda,k})^\top \mathbf{R}_{\lambda.k}^{-1}(\mathbf{y}_k-\boldsymbol{\mu}_{\lambda,k})=\frac{1}{g} \mathbf{m}_k^T \left(\mathbf{I}_{n_v}+\frac{1}{n_t+g}\mathbf{J}_{n_v}\right)\mathbf{m}_k\ ,\ \]
where $\mathbf{m}_k=(\mathbf{I}_{n_v}-\frac{1}{n+g}\mathbf{J})(\mathbf{y}_k - \frac{n_t\bar{y}_{-k}}{n_t+g}\mathbf{1}_{n_v})$. It is straightforward to show $\mathbf{m}_k=\mathbf{y}_k-\tilde{y}\mathbf{1}_{n_v}$ where $\tilde{y}=\frac{n}{n+g}\overline{y}=0$ due to centering the data. Hence
\begin{align*}
\lim_{\lambda \to \infty} (\mathbf{y}_k-\boldsymbol{\mu}_{\lambda,k})^\top \mathbf{R}_{\lambda.k}^{-1}(\mathbf{y}_k-\boldsymbol{\mu}_{\lambda,k})&=\frac{1}{g}\mathbf{y}_k^T\left(\mathbf{I}_{n_v}+\frac{1}{n_t+g}\mathbf{J}_{n_v}\right)\mathbf{y}_k\\
&=\frac{1}{g}\left(\mathbf{y}_k^T\mathbf{y}_k+\frac{n_v^2}{n_t+g}\bar{y}_k^2\right)\ ,\
\end{align*}
and we arrive at the final expression
\begin{align*}
    \lim_{\lambda \to \infty} \text{MD}(\mathcal{D}_k \mid \mathcal{D}_{-k}, \lambda)&=n_t \ \frac{\mathbf{y}_k^T\mathbf{y}_k+\frac{n_v^2}{n_t+g}\bar{y}_k^2}{\mathbf{y}_{-k}^T\left(\mathbf{I}_{n_t}-\frac{1}{n_t+g}\mathbf{J}_{n_t}\right)\mathbf{y}_{-k}}\\
    &=n_t \ \frac{\mathbf{y}_k^T\mathbf{y}_k+\frac{n_v^2}{n_t+g}\bar{y}_k^2}{\mathbf{y}_{-k}^T\mathbf{y}_{-k}-\frac{n_t^2}{n_t+g}\bar{y}_{-k}^2}\\
    &\approx n_t \ \frac{\mathbf{y}_k^T\mathbf{y}_k+\frac{n_v^2}{n_t}\bar{y}_k^2}{\mathbf{y}_{-k}^T\mathbf{y}_{-k}-n_t\bar{y}_{-k}^2}\ .\
\end{align*}

    \subsection{Limiting form of Score}
    Score is defined as 
    \begin{equation*}
        \begin{aligned}
            \textrm{Score}(\mathcal{D}_k \mid \mathcal{D}_{-k}, \lambda)&=\textrm{MD}(\mathcal{D}_k \mid \mathcal{D}_{-k}, \lambda)+\log|\boldsymbol{\Sigma}_{\lambda,k}|\\
            &=\textrm{MD}(\mathcal{D}_k \mid \mathcal{D}_{-k}, \lambda)+ n_v \log(\hat{\sigma}^2_{\lambda,-k}) + \log|\mathbf{R}_{\lambda,k}|\ .\
        \end{aligned}
    \end{equation*}
    As $\lambda\to \infty$, the variance term satisfies 
    \begin{equation*}
        \lim_{\lambda \to \infty}n_v \log(\hat{\sigma}^2_{\lambda,-k})=n_v\left\{-\log(g)-\log(n_t)+\log \left(\mathbf{y}^\top_{-k}\left(\mathbf{I}_{n_t}-\frac{1}{n_T+g}\mathbf{J}_{n_t}\right)\mathbf{y}^\top_{-k}\right)\right\} \ .\
    \end{equation*}
   For the determinant term, we consider eigenvalues of $\mathbf{R}_{\lambda,k}$. As $\theta\to 0$, $\mathbf{R}_{\lambda,k}=g\left(\frac{1}{g+n_t}\mathbf{J}_{n_v} + \mathbf{I}_{n_v}\right)$. Since $\mathbf{J}_{n_v}$ has eigenvalues $n_v$ (once) and 0 (multiplicity $n_v-1$), the eigenvalues of $\mathbf{R}_{\lambda,k}$ are $g\left(1+\frac{n_v}{g+n_t}\right)$ (once) and $g$ (multiplicity $n_v-1$). Therefore, 
    \begin{equation*}
        \log|\mathbf{R}_{\lambda,k}| = \log \left(1+\frac{n_v}{g+n_t}\right)+  n_v\log (g)
    \end{equation*}
    Combining terms, as $\lambda\to \infty$,
\begin{align*}
     \lim_{\lambda \to \infty}\textrm{Score}(\mathcal{D}_k \mid \mathcal{D}_{-k}, \lambda)&=\lim_{\lambda \to \infty}\textrm{MD}(\mathcal{D}_k \mid \mathcal{D}_{-k}, \lambda) - n_v\log(n-n_v) + n_v\log(\textrm{SS}_{-k}) + \log \left(1+\frac{n_v}{g+n_t}\right) \\
     &\approx  n_t \ \frac{\mathbf{y}_k^T\mathbf{y}_k+\frac{n_v^2}{n_t}\bar{y}_k^2}{\mathbf{y}_{-k}^T\mathbf{y}_{-k}-n_t\bar{y}_{-k}^2} - n_v\log(n_t) + n_v\log \left(\mathbf{y}_{-k}^T\mathbf{y}_{-k}-n_t\bar{y}_{-k}^2\right) + \log\left(1+\frac{n_v}{n_t}\right)
\end{align*}
Here, the approximation $g+n_t\approx n_t$ holds since $g$ is set to be small.

\subsection{Limiting form of PE}
We can also show the limit of $\text{PE}_k$ as $\lambda\to\infty$ is
\[\lim_{\lambda \to \infty} \text{PE}_k = \left(\mathbf{y}_k - \frac{n_t\bar{y}_{-k}}{n_t+g}\mathbf{1}_{n_v}\right)^T\left(\mathbf{y}_k - \frac{n_t\bar{y}_{-k}}{n_t+g}\mathbf{1}_{n_v}\right)\ , \]
which becomes small when the $\mathbf{y}_k$ lie close to the training data mean $\bar{y}_{-k}$. This behavior can be remedied by ensuring that each fold adequately fills the output space.  In the case of $n$-fold CV, $\bar{y}_{-k}=(n-1)^{-1}(n\bar{y}-y_k)=-(n-1)^{-1}y_k$ since $\bar{y}=0$, since we center data. Substituting into the expression gives
\[\lim_{\lambda \to \infty} \text{PE}_k = \left(y_k + y_k\frac{n-1}{n-1+g}\frac{1}{n-1}\right)^2=y_k^2\left(1+\frac{1}{n-1+g}\right)^2= y_k^2 \left(\frac{n+g}{n-1+g}\right)^2 .\]
Thus 
\[\lim_{\lambda \to \infty}\text{PE}(\lambda)\propto \frac{1}{n}\sum y_i^2 \ ,\]
which provides a desirable upper bound. Unlike MD, this ensures that $n$-fold CV based on PE avoids selecting $\lambda$ whose surrogate fail to interpolate the data.

\subsection{Limiting forms of each metric for large lengthscale parameters}
It is also informative to examine the opposite extreme - flat likelihoods that make $\boldsymbol{\theta}\to \infty$. For large values of  $\theta$, if $x_i\ne x_j$, then $|x_i-x_j|>0$, and thus $R(x_i, x_j)=\mathrm{exp}(-\theta(x_i-x_j)^2)\to 0$ as $\theta\to \infty$. When $x_i=x_j$, we have $R(x_i, x_i)=1$ for all $\theta$. Consequently, as $\theta\to \infty$, $\mathbf{R}(\mathcal{X}, \mathbf{X}_n)\approx \mathbf{0}$, and $\mathbf{R}_n\approx \mathbf{I}$. It follows that 
\begin{equation*}
    \begin{aligned}
        \boldsymbol{\mu}_\theta(\mathcal{X} \mid \mathcal{D}) & = \mathbf{R}(\mathcal{X}, \mathbf{X}_n) \mathbf{R}_n^{-1}\mathbf{y}_n \\
    &=\mathbf{0}\\
     \mathbf{\Sigma}_\Omega(\mathcal{X} \mid \mathcal{D}) &= \hat{\sigma}_\theta^2\left( \mathbf{R}(\mathcal{X},\mathcal{X})- \mathbf{R}(\mathcal{X}, \mathbf{X}_n)\mathbf{R}_n^{-1} \mathbf{R}(\mathbf{X}_n,\mathcal{X})\right) \\
     &=\hat{\sigma}_\theta^2\mathbf{I} \ . 
    \end{aligned}
\end{equation*}
In this case, the correlation matrix approaches the identity, and each metric reduces to a simple closed form:
\begin{align*}
    \lim_{\theta \to \infty} \text{PE}_k &= \mathbf{y}_k^T\mathbf{y}_k\\
    \lim_{\theta \to \infty} \text{MD}_k &= n_t \ \frac{\mathbf{y}_k^T\mathbf{y}_k}{\mathbf{y}_{-k}^T\mathbf{y}_{-k}}\\
    \lim_{\theta \to \infty} \text{DPE}_k &= \frac{1}{1+g} \mathbf{y}_k^T\mathbf{y}_k\approx \mathbf{y}_k^T\mathbf{y}_k\ ,\
\end{align*}
where the approximation holds since $g$ is taken to be small.

\section{Test Functions}
\paragraph{Lim Nonpolynomial.} The two-dimensional Lim Nonpolynomial function \citep{limfunction} is defined as 
\begin{equation*}
    f(\mathbf{x})=\frac{1}{6}[(30+5x_1\sin{(5x_1)})(4+\exp{(-5x_2)})-100] \quad  x_i\in[0,1]
\end{equation*}

\paragraph{Franke.} The two-dimensional Franke function \citep{frankefunction} is defined as 
\begin{equation*}
    \begin{aligned}
        f(\mathbf{x})=& \; 0.75 \exp\!\left(-\frac{(9x_1 - 2)^2}{4} - \frac{(9x_2 - 2)^2}{4}\right) \\
    &+ 0.75 \exp\!\left(-\frac{(9x_1 + 1)^2}{49} - \frac{9x_2 + 1}{10}\right) \\
    &+ 0.5 \exp\!\left(-\frac{(9x_1 - 7)^2}{4} - \frac{(9x_2 - 3)^2}{4}\right) \\
    &- 0.2 \exp\!\left(-(9x_1 - 4)^2 - (9x_2 - 7)^2\right), 
    \quad x_i \in [0,1].
    \end{aligned}
\end{equation*}

\paragraph{Piston Simulation.} The seven-dimensional Piston Simulation function \citep{pistonfunction1} is defined as 
\begin{equation*}
    f(\mathbf{x})=2\pi\sqrt{\frac{M}{k+S^2\frac{P_0V_0}{T_0}\frac{T_a}{V^2}}} \ ,
\end{equation*}
where 
\begin{equation*}
     V=\frac{S}{2k}\left(\sqrt{A^2+4k\frac{P_0V_0}{T_0}T_a-A}\right), \quad A=P_0S+19.62M-\frac{kV_0}{S}.
\end{equation*}
The input ranges are given by  $M\in[30, 60]$, $S\in[0.005, 0.020]$, $V_0\in [0.002, 0.010]$, $k\in[1000, 5000]$, $P_0\in[90000, 110000]$, $T_a\in[290, 296]$, $T_0\in [340, 360]$

\paragraph{Borehole.} The eight-dimensional Borehole function \citep{borehole} is defined as 
\begin{equation*}
    f(\mathbf{x})=\frac{2\pi T_u(H_u-H_l)}{\log(r/r_w)\left(1+\frac{2LT_u}{\log(r/r_w)r^2_wK_w}\right)+\frac{T_u}{T_l}} .
\end{equation*}
The input ranges are given by $r_2\in[0.05, 0.15]$, $r\in [100, 50000]$, $T_u\in[63070, 115600]$, $H_u\in[990, 1110]$, $T_l\in[63.1, 116]$, $H_l\in [700, 820]$, $L\in[1120, 1680]$, $K_w\in[9855, 12045]$.

\section{Piston Slap Noise Data}\label{app:piston}
\begin{table}[H]
    \centering
    \begin{tabular}{|ccccccc|}
    \hline
        $x_1$ & $x_2$ & $x_3$ & $x_4$ & $x_5$ & $x_6$ & Noise(dB)\\ 
        \hline
        71 &  16.8 & 21.0 & 2 & 1 & 0.98 & 56.75 \\
        \hline
                 15 & 15.6 & 21.8 & 1 & 2 & 1.30 & 57.65 \\
                 \hline
                 29 & 14.4 & 25.0 & 2 & 1 & 1.14 & 53.97 \\
                 \hline
                 85 & 14.4 & 21.8 & 2 & 3 & 0.66 & 58.77 \\
                 \hline
                 29 & 12.0 & 21.0 & 3 & 2 & 0.82 & 56.34 \\
                 \hline
                 57 & 12.0 & 23.4 & 1 & 3 & 0.98 & 56.85 \\
                 \hline
                 85 & 13.2 & 24.2 & 3 & 2 & 1.30 & 56.68 \\
                 \hline
                 71& 18.0 & 25.0& 1& 2& 0.82& 58.45\\
                 \hline
                 43& 18.0& 22.6& 3& 3& 1.14& 55.50\\
                 \hline
                 15& 16.8& 24.2& 2& 3& 0.50& 52.77\\
                 \hline
                 43& 13.2& 22.6& 1& 1& 0.50& 57.36\\
                 \hline
                 57& 15.6& 23.4& 3& 1& 0.66& 59.64\\
                 \hline
    \end{tabular}
    \caption{Training data for piston noise slap example in \Cref{sec:piston}.}
\end{table}

\section{Piston Slap Noise Data: Anisotropic kernel} \label{app:anisotropic}
We also fit our method using an anisotropic kernel. We first evaluated the performance of MLE, resulting in an RMSE of 0.767 and a CRPS of 0.500, and $n$-fold CV with PE, resulting in an RMSE of 0.768 and a CRPS of 0.504 (a slightly inferior result).

\begin{figure}[!ht]
    \centering
     \subfloat[RMSE\label{fig:piston_rmse_appendix}]{\includegraphics[width=0.38\linewidth]{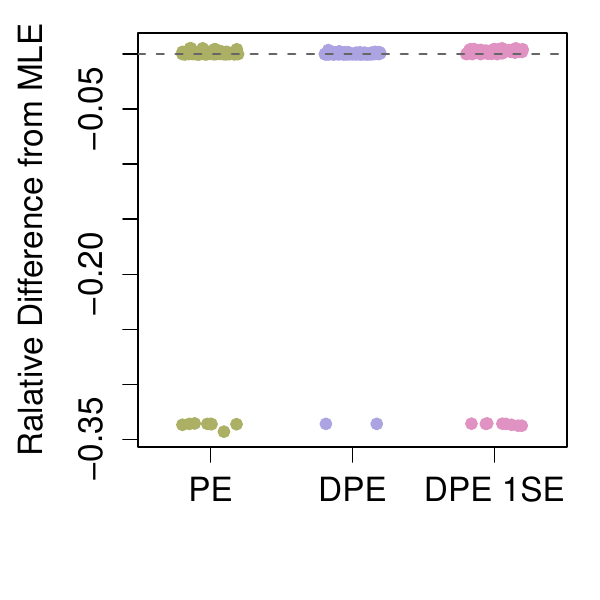}}
     \subfloat[CRPS\label{fig:piston_crps_appendix}]{\includegraphics[width=0.38\linewidth]{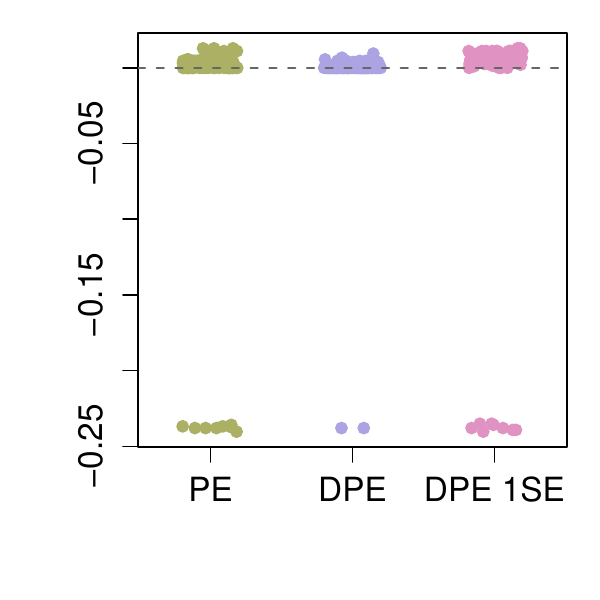}}

    \caption{Relative differences in RMSE and CRPS compared with the MLE baseline. The gray dashed line represents zero difference, corresponding to selecting $\lambda=0$.}
    \label{fig:piston_anisotropic}
\end{figure}

We then implemented $4$-fold CV with PE, DPE, and DPE 1SE. Since the performance of $4$-fold CV can depend on the construction of the folds, we repeated the procedure for 100 randomly generated folds. We evaluated performance by comparing RMSE/CRPS against the ``benchmark'' MLE values. \Cref{fig:piston_anisotropic} shows the relative differences in RMSE and CRPS for all 100 repetitions. A value of zero indicates that each method selected selected no penalization ($\lambda=0$), positive values correspond to worse performance than MLE, and negative values indicate improvement. As the figure shows, even though each metric occasionally performs worse than MLE, the degradation is generally small. PE and DPE 1SE achieve improvements over MLE more frequently than DPE, and when they do, they are able to achieve the significant improvement. These results highlight the potential benefits of penalization in limited sample settings. However, they also suggest that the sensitivity of $4$-fold CV to fold configuration. The folds were generated completely at random in this experiment. A more structured fold design may yield more reliable results than random partitioning.

\begin{table}[H]
    \centering
    \begin{tabular}{|c||c|c|c|}
                \hline
                    & MLE & $n$-fold  PE&  Best $4$-fold CV   \\
                    \hline
                 $\lambda$  & 0& 0.006 & 0.058   \\
                 \hline
                 $\hat{\sigma}^2_\theta$ & 1.151  & 1.241 & 5.382 \\
                 \hline
                 $\hat{\theta}_1$ & 4.067 & 3.728 & 0.387    \\
                 \hline 
                 $\hat{\theta}_2$ & 0.001 & 0.001 & 0.001\\
                 \hline
                 $\hat{\theta}_3$ & 0.588 & 0.532 & 0.001 \\
                 \hline
                 $\hat{\theta}_4$& 0.001  & 0.001 & 0.906 \\
                 \hline 
                 $\hat{\theta}_5$& 0.001  & 0.001 &0.019\\
                 \hline 
                 $\hat{\theta}_6$& 2.751  & 2.550 & 0.428 \\
                 \hline
                \end{tabular}
    \caption{Parameter estimates from the piston slap noise data for MLE, $n$-fold CV with PE, and a $4$-fold CV providing improved RMSE/CRPS.
    }
    \label{tab:piston_estimates_anisotopic}
\end{table}
To provide more context for this improved performance, \Cref{tab:piston_estimates_anisotopic} shows $\hat{\sigma}_\theta^2$ and $\hat{\theta}_p$ for the MLE and $n$-fold CV with PE surrogates alongside the estimates for $\lambda=0.058$, which produced the median RMSE value of $0.431$ for the 4-fold CV metrics. Even slight penalization with this $\lambda$ led to distinctly different estimates. Notably, $\hat{\theta}_1$ and $\hat{\theta}_6$ were shrunk significantly, yet $\hat{\theta}_2$ was unaffected by the penalization.  Perhaps the most impactful changes occurred for $\hat{\theta}_4$, which went from the smallest lengthscale under MLE and $n$-fold CV with PE to the largest lengthscale under the penalization. These findings illustrate that even slight penalization can reshape hyperparameter estimates and influence model behavior and performance. 

\end{document}